\documentclass[journal]{IEEEtran}
\usepackage[utf8]{inputenc}
\usepackage[T1]{fontenc}
\usepackage{mathptmx}
\usepackage{microtype}
\usepackage{indentfirst}
\usepackage{amsmath, amssymb, bm}
\usepackage{amsthm}
\usepackage{upgreek}
\usepackage{algorithm}
\usepackage{algpseudocode}

\usepackage{graphicx}
\usepackage{tikz}
\usetikzlibrary{positioning, arrows.meta, shapes.geometric, arrows, fit}
\usepackage{pgfplots}
\usepackage{pgffor}
\pgfplotsset{compat=1.18}
\usepackage{booktabs}
\usepackage{multirow}
\usepackage{array, tabularx, colortbl}
\usepackage{float}
\usepackage{caption}
\usepackage{setspace}
\usepackage{titlesec}
\usepackage{enumitem}
\usepackage{cite}
\usepackage{hyperref}
\usepackage{cleveref}
\usepackage{xcolor}
\usepackage{soul}
\usepackage{tcolorbox}
\tcbuselibrary{fitting}
\usepackage{mdframed}
\usepackage{url}
\usepackage{lastpage}
\usepackage{etoolbox}
\usepackage{lineno}
\usepackage{changepage}
\usepackage{ragged2e}
\usepackage{footmisc}
\usepackage{marginnote, marginfix}
\usepackage{scrextend}
\usepackage{pbox}
\usepackage{neuralnetwork}
\definecolor{swarmblue}{RGB}{30,144,255}
\definecolor{coopergreen}{RGB}{34,139,34}
\definecolor{neuralred}{RGB}{220,20,60}

\title{SwarmSense-DNN: A Trustworthy and Decentralized Neural Framework for Proactive Anomaly Defense in Consumer IoT}

\author{Jing Yang,
        Vijay Govindarajan,
        Saad Arif,
        Xu Xu*,
        Mohamed Kallel,
        Zaffar Ahmed Shaikh,
        Zhe Liu,
        Chunhong Yuan,
        Lip Yee Por*
\thanks{This work was supported by the Deanship of Scientific Research, Vice Presidency for Graduate Studies and Scientific Research, King Faisal University, Saudi Arabia [Grant No. KFU253799].}
\thanks{Corresponding author: Lip Yee Por and Xu Xu}
\thanks{Jing Yang and Por Lip Yee are with the Center of Research for Cyber Security and Network (CSNET), Faculty of Computer Science and Information Technology, Universiti Malaya, 50603 Kuala Lumpur, Malaysia (e-mail: s2147529@siswa.um.edu.my; porlip@um.edu.my).}
\thanks{Vijay Govindarajan is with the Distribution and Supply Technology, Expedia Group, Seattle, WA 98119, United States (e-mail: vigovindaraja@expediagroup.com).}
\thanks{Saad Arif is with the Department of Mechanical Engineering, College of Engineering, King Faisal University, Al Ahsa, 31982, Saudi Arabia. (e-mail: sarif@kfu.edu.sa).}
\thanks{Xu Xu is with the School of Computer Science and Engineering, Northeastern University, Shenyang 110004, China (e-mail: xuxu@ieee.org).}
\thanks{Mohamed Kallel is with the Department of Physics, College of Science, Northern Border University, Arar, Saudi Arabia. (e-mail: Mohamed.Kallel@nbu.edu.sa).}
\thanks{Zaffar Ahmed Shaikh is with Department of Computer Science and Information Technology, Benazir Bhutto Shaheed University Lyari, Karachi, 75660, Pakistan and with School of Engineering, \'Ecole Polytechnique F\'ed\'erale de Lausanne, 1015, Lausanne, Switzerland (zashaikh@bbsul.edu.pk).}
\thanks{Zhe Liu is with College of Mathematics and Computer, Xinyu University, Xinyu 338004, China and with School of Computer Sciences, Universiti Sains Malaysia, Penang 11800, Malaysia (zheliu@ieee.org).}
\thanks{Chunhong Yuan is with Faculty of Control Systems and Robotics, National Research University for Information Technology, Mechanics and Optics (ITMO), St. Petersburg, 197101, Russia (Y2399549081@outlook.com).}
}

\begin{document}
\maketitle

\begin{abstract}
The rapid growth of consumer IoT devices has introduced unprecedented challenges in trustworthy anomaly detection against AI-enabled cyber threats, requiring real-time, privacy-preserving, and scalable defense mechanisms. Traditional centralized strategies face critical limitations, including communication bottlenecks, single points of failure, and privacy vulnerabilities when processing distributed consumer data. We propose SwarmSense-DNN, a novel decentralized neural framework employing swarm intelligence for secure, cooperative anomaly detection across distributed IoT environments. The framework integrates autonomous agents with deep neural networks to form a self-organizing defense system that detects evolving anomalies without centralized coordination. It utilizes hierarchical federated learning with graph neural networks and attention mechanisms to capture local and global anomaly behaviors while ensuring data privacy. Extensive experiments demonstrate SwarmSense-DNN's superior performance: it achieves 95.44\% average detection accuracy across five benchmark datasets while reducing communication overhead by 67\%. The framework maintains robust resilience against adversarial threats through differential privacy safeguards and demonstrates strong fault tolerance under node failures and AI-enabled attacks.
\end{abstract}


\section{Introduction}

The rapid growth of interconnected systems across smart homes, consumer devices, and industrial IoT~\cite{lin2023pushing,qu2025mobile,fang2025dynamic} has reshaped trustworthy anomaly detection in distributed environments increasingly exposed to AI-enabled cyber threats~\cite{Mothukuri2022, bao2026sider,aleisa2023transforming, yang2025neuroagent,zhang2025state}. Centralized detectors can work in controlled settings but face key limits in consumer networks where privacy, bandwidth, and real-time processing are essential~\cite{yang2025llm, yang2026neurothyroid,lin2025hierarchical,hong2026conflict,sun2025rrto,fang2024automated,zhang2024satfed,lin2025hsplitlora}. These constraints are acute in consumer and critical infrastructure scenarios, where anomalies may signal AI-driven intrusions, device faults, or operational risks that demand immediate response~\cite{zeng2024tensor, yang2025wastewater}. The spread of edge computing and ubiquitous IoT has therefore pushed a shift toward decentralized, on-device processing that supports localized detection and decision-making~\cite{Liu2022, dai2024intrusion, yang2026neurocare,lin2024adaptsfl}. Yet many distributed methods still rely on simplified aggregation that fails to capture complex temporal dynamics and internode dependencies in heterogeneous systems~\cite{Kumari2024, yang2026dfgnet,lin2026gapsl}. The diversity of data sources, together with strict privacy rules and bandwidth limits, calls for trustworthy frameworks that maintain high accuracy and efficiency under edge constraints~\cite{Dong2025, gupta2023artificial, yang2025towards}.

Swarm intelligence, inspired by the collective behavior of biological entities such as ant colonies and bee swarms, offers a promising foundation for building decentralized and adaptive defense mechanisms in distributed systems~\cite{Wardhana2025, cheng2024q, yang2025swarm,lin2025hasfl}. Unlike traditional distributed systems that rely on centralized coordination, swarm-based approaches enable autonomous agents to collaborate through local interactions, emerging complex global behaviors without explicit central control~\cite{Pham2021, yang2025optimized}. This self-organizing coordination mechanism aligns naturally with the goals of trustworthy distributed anomaly detection systems, enabling nodes to make local defense decisions while contributing to a unified and system-wide threat perspective~\cite{Blais2023, yang2026explainable}.

Recent developments in federated learning have shown the feasibility of training models across decentralized datasets while preserving data privacy, which is an essential requirement for secure consumer applications~\cite{Man2021, popoola2023federated, khan2024asmf, yang2025enhancing}. However, many federated learning methods for anomaly detection still rely solely on model aggregation, overlooking the potential of swarm intelligence to enhance robustness, adaptability, and cyber threat detection accuracy~\cite{Li2022, fahim2025resource, yang2025qb}. The integration of swarm intelligence with deep neural networks presents an opportunity to develop scalable, adaptive, and trustworthy anomaly detection frameworks tailored for dynamic and adversarial consumer IoT environments~\cite{Syu2023, yang2025ddos}.

Modern consumer IoT faces challenges from privacy constraints, limited communication, and growing demands for scalability and reliability. Centralized architectures cause single points of failure, bandwidth inefficiency, and privacy risks~\cite{khan2025baiot, yang2026blockchain}. Conventional federated learning with simple aggregation fails to capture temporal and cross-node dependencies in heterogeneous networks. The key goal is to build a decentralized and trustworthy framework that achieves high detection accuracy with privacy preservation and low communication cost, while adapting to dynamic environments, resisting AI-enabled adversaries, and tolerating node failures under resource constraints~\cite{Liu2023, yang2026cardiotwin}.

This work advances IoT security through SwarmSense-DNN's novel integration of swarm intelligence with federated learning, enabling distributed anomaly detection without centralized data aggregation, a departure from traditional centralized security models. SwarmSense-DNN introduces a decentralized paradigm where consumer IoT devices collaboratively detect anomalies through pheromone-inspired coordination and local model training, eliminating the privacy risks inherent in centralized architectures. This paper makes the following contributions toward secure and decentralized anomaly detection in consumer IoT.

\begin{enumerate}
\item We propose \textbf{SwarmSense DNN}, a trustworthy decentralized framework that fuses swarm intelligence with deep neural networks for collaborative, proactive anomaly detection in consumer IoT~\cite{yang2025swarm}. Autonomous nodes learn and adapt collectively without relying on centralized coordination.
\item We design a hierarchical architecture that integrates local swarm clusters with lightweight coordination mechanisms to balance detection accuracy, scalability, and communication efficiency. The architecture employs adaptive clustering and selective information exchange to reduce overhead.
\item We introduce differential privacy mechanisms tailored for swarm-based learning. These techniques protect individual node data while enabling effective collaboration across distributed systems.
\item We implement self-healing strategies that autonomously detect and recover from node failures or compromises. These mechanisms enhance system reliability, trustworthiness, and resilience in adversarial and dynamic conditions.
\end{enumerate}

\section{Literature Review}

\subsection{Distributed and Federated Anomaly Detection}
Federated learning is a leading approach for privacy-preserving distributed ML, with broad uses in anomaly detection~\cite{Dong2025}. Mothukuri et al.~\cite{Mothukuri2022} validated its feasibility for IoT security, showing that accuracy can be maintained while protecting data, though simple parameter averaging struggles with complex, heterogeneous anomaly patterns. Recent work tackles heterogeneity, communication, and decentralization. Man et al.~\cite{Man2021} used adaptive aggregation in edge-assisted IoT to handle data quality variation and reduce latency but relied on centralized coordination. Li et al.~\cite{Li2022} advanced decentralized FL to remove single points of failure and improve resilience, yet lacked adaptive coordination for evolving threats. Zhang et al.~\cite{Zhang2025} introduced a spatially aware GNN that captures topology while keeping a centralized setup that limits scalability and privacy. Syu et al.~\cite{Syu2023} studied privacy-aware energy grids, clarifying privacy--accuracy trade-offs in critical infrastructure.

\subsection{Swarm Intelligence and Cooperative Systems}
Swarm intelligence is a powerful approach for optimization and coordination in distributed systems~\cite{Wardhana2025}. Wardhana et al. surveyed core concepts and applications, outlining foundations for integrating machine learning. Xu et al.~\cite{Xu2023} introduced cooperative swarm learning with cyclic updates that cut communication while preserving effectiveness, though it was not aimed at anomaly detection. Pham et al.~\cite{Pham2021} showed how bio-inspired methods tackle large-scale, dynamic network challenges, offering insights for swarm-based anomaly detection. In robotics, Blais and Akhloufi~\cite{Blais2023} reviewed reinforcement learning for swarm coordination and demonstrated the value of decentralized decision-making, which we adapt for anomaly detection in distributed systems. In clinical contexts, Wardhana et al.~\cite{Wardhana2021} validated decentralized, privacy-preserving collaboration, directly informing our confidentiality mechanisms.

\subsection{Deep Learning Architectures for Anomaly Detection}
Deep learning for anomaly detection has progressed rapidly. Surveys by Huang et al.~\cite{Huang2025} and Kumari et al.~\cite{Kumari2024} chart the evolution of neural architectures and explain how designs are tuned for different anomaly types and deployment contexts. Building on this foundation, Nazir et al.~\cite{Nazir2024} introduce hybrid CNN--LSTM models for IoT threat detection that capture spatial--temporal patterns under resource constraints; Singh et al.~\cite{Singh2025} boost accuracy with hybrid deep models augmented by optimization and ensemble techniques; and Chen et al.~\cite{Chen2025} integrate fuzzy clustering--based signature intrusion detection with deep learning to demonstrate the benefits of combining traditional and modern methods.

\subsection{Graph Neural Networks and Attention Mechanisms}
Graph neural networks are central to anomaly detection in networked systems. Kim et al.~\cite{Kim2022} reviewed methods and challenges, outlining foundations and limits. Zhou et al.~\cite{Zhou2024} introduced grouped GNNs for multivariate time series, capturing dependencies across sensors and groups. Tang et al.~\cite{Tang2022} rethought GNN architectures for anomaly detection, offering insights that guide our swarm-based coordination. Dadhania et al.~\cite{Dadhania2025} applied GNNs to software-defined networks for high-speed anomaly detection, showing scalability in real-time settings. Chen et al.~\cite{Chen2025} advanced transformer models with improved attention for time series anomalies, capturing long-range temporal patterns. Liu et al.~\cite{Liu2025} used inter-variable attention for multivariate anomalies, informing our attention-driven swarm coordination. Kavadi et al.~\cite{Kavadi2025} combined temporal graph attention with transformer-augmented RNNs, highlighting the promise of hybrid attention architectures.

\subsection{Edge Computing and IoT Applications}
Edge anomaly detection has been widely studied under tight latency, bandwidth, and energy limits. Liu et al.~\cite{Liu2022} outlined deployment principles for resource-constrained industrial IoT, while Rezaee et al.~\cite{Rezaee2022} showed the need for real-time distributed pipelines in edge video surveillance. Practical systems include IoT intrusion detection that balances accuracy with resources~\cite{Mudgerikar2020} and frameworks with dynamic node insertion and deletion for evolving streams and topologies~\cite{Xiang2022}. Domain-specific designs such as multi-sensor detection in underground mining further illustrate edge-centric constraints and choices~\cite{Zhuang2021}.

\subsection{Ensemble Learning and Variational Methods}
Ensemble learning has shown strong potential for enhancing anomaly detection robustness and accuracy. Sarhan et al.~\cite{Sarhan2020} developed deep ensembles for detecting network anomalies and cyber attacks, proving that combining multiple models improves detection performance over single approaches. Liu et al.~\cite{Liu2024} introduced selective parallel ensemble methods that optimize the balance between diversity and accuracy, offering design insights for distributed and scalable systems. Variational autoencoders have gained prominence in anomaly detection, with Nguyen et al.~\cite{Nguyen2024} providing a comprehensive comparison that establishes benchmarks and architectural guidelines. Wang et al.~\cite{Wang2025} integrated LSTM-based autoencoders with attention to detect multidimensional temporal anomalies, highlighting the benefit of combining recurrent and attention mechanisms. Ji et al.~\cite{Ji2025} further demonstrated the scalability of VAE-based unsupervised models in large-scale energy storage systems, illustrating their adaptability to critical infrastructure and real-world deployment scenarios.

\subsection{Research Gap Analysis}
Although distributed anomaly detection has advanced, a key gap remains in integrating swarm intelligence with deep learning for autonomous cooperative systems. Existing methods rely on simple FL aggregation~\cite{Mothukuri2022, Man2021,lin2024fedsn} or centralized neural models~\cite{Zhang2025, Chen2025}, lacking adaptive coordination, privacy, efficiency, and resilience~\cite{Jokic2025}. No work has yet combined swarm-based coordination~\cite{Pham2021, Xu2023} with powerful neural architectures to build self-organizing, privacy-preserving anomaly detection adaptable to evolving threats~\cite{Dong2025, Liu2023}. This gap motivates our SwarmSense-DNN framework. While recent federated approaches like Dong et al.'s FADngs~\cite{Dong2025} and Popoola et al.'s consumer IoT intrusion detection~\cite{popoola2023federated} achieve distributed learning, they maintain centralized coordination mechanisms. Xu et al.'s cooperative swarm learning~\cite{Xu2023} demonstrates bio-inspired coordination but lacks trust-based consensus, while Zhang et al.'s GNN approach~\cite{Zhang2025} uses static network topologies rather than adaptive clustering. Current hierarchical methods like Li et al.'s decentralized FL~\cite{Li2022} rely on simple parameter averaging instead of bio-inspired collective decision-making, leaving a fundamental gap in truly autonomous distributed systems that can self-organize without centralized control.

\section{Methodology}

\subsection{System Model and Assumptions}

\subsubsection{System Model}
We consider a distributed network comprising $N$ autonomous nodes $\mathcal{N} = \{n_1, n_2, \ldots, n_N\}$ deployed across a heterogeneous environment. The network topology is modeled as a dynamic graph $\mathcal{G}(t) = (\mathcal{N}, \mathcal{E}(t), \mathcal{W}(t))$, where $\mathcal{E}(t) \subseteq \mathcal{N} \times \mathcal{N}$ represents time-varying communication links and $\mathcal{W}(t) \in \mathbb{R}^{N \times N}$ denotes the weighted adjacency matrix. The connectivity between nodes $i$ and $j$ at time $t$ is defined as:
\begin{equation}
W_{ij}(t) = \begin{cases}
\exp(-d_{ij}/\sigma_c) & \text{if } d_{ij} \leq r_c \\
0 & \text{otherwise}
\end{cases}
\end{equation}
where $d_{ij}$ is the distance between nodes, $r_c$ is the communication range, and $\sigma_c$ controls the connectivity strength decay.

Each node $n_i$ continuously observes local data streams $\mathbf{X}_i(t) = [\mathbf{x}_i(t-T+1), \ldots, \mathbf{x}_i(t)] \in \mathbb{R}^{d_i \times T}$ over sliding time windows of length $T$. The local feature vector at time $t$ is denoted as $\mathbf{x}_i(t) \in \mathbb{R}^{d_i}$, where $d_i$ represents the dimensionality of observations at node $i$. We assume that the data distribution at each node follows:
\begin{equation}
\mathbf{x}_i(t) \sim \mathcal{D}_i(t) = (1-\lambda_i)\mathcal{D}_{\text{normal}} + \lambda_i\mathcal{D}_{\text{anomaly}}
\end{equation}
where $\lambda_i \in [0,1]$ represents the anomaly rate at node $i$, $\mathcal{D}_{\text{normal}}$ is the normal data distribution, and $\mathcal{D}_{\text{anomaly}}$ represents the anomalous distribution.

\begin{figure}[!htb]
\centering
\includegraphics[width=0.8\columnwidth]{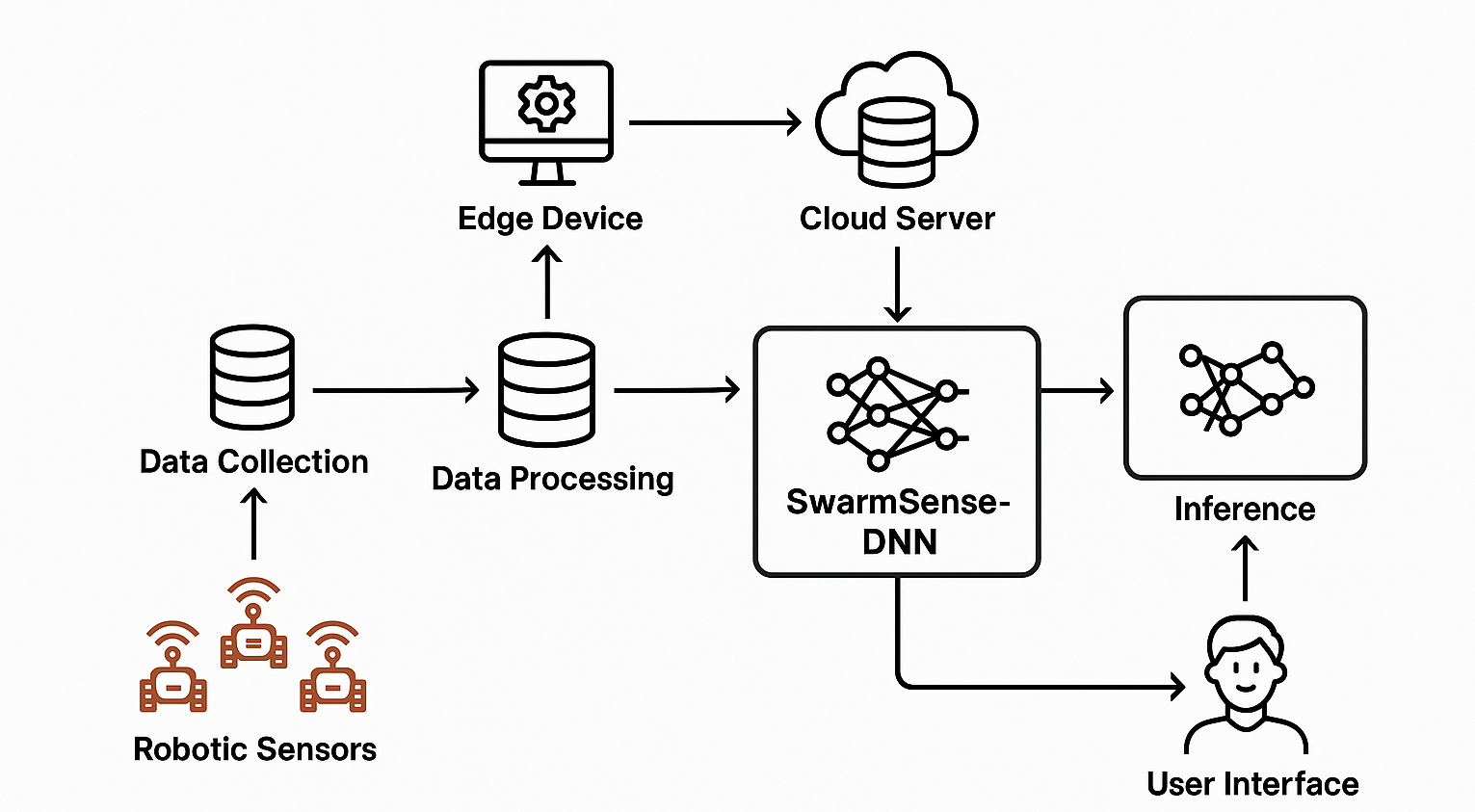}
\caption{SwarmSense-DNN system model.}
\label{fig:system_model}
\end{figure}

\begin{figure}[!htb]
\centering
\includegraphics[width=0.8\columnwidth]{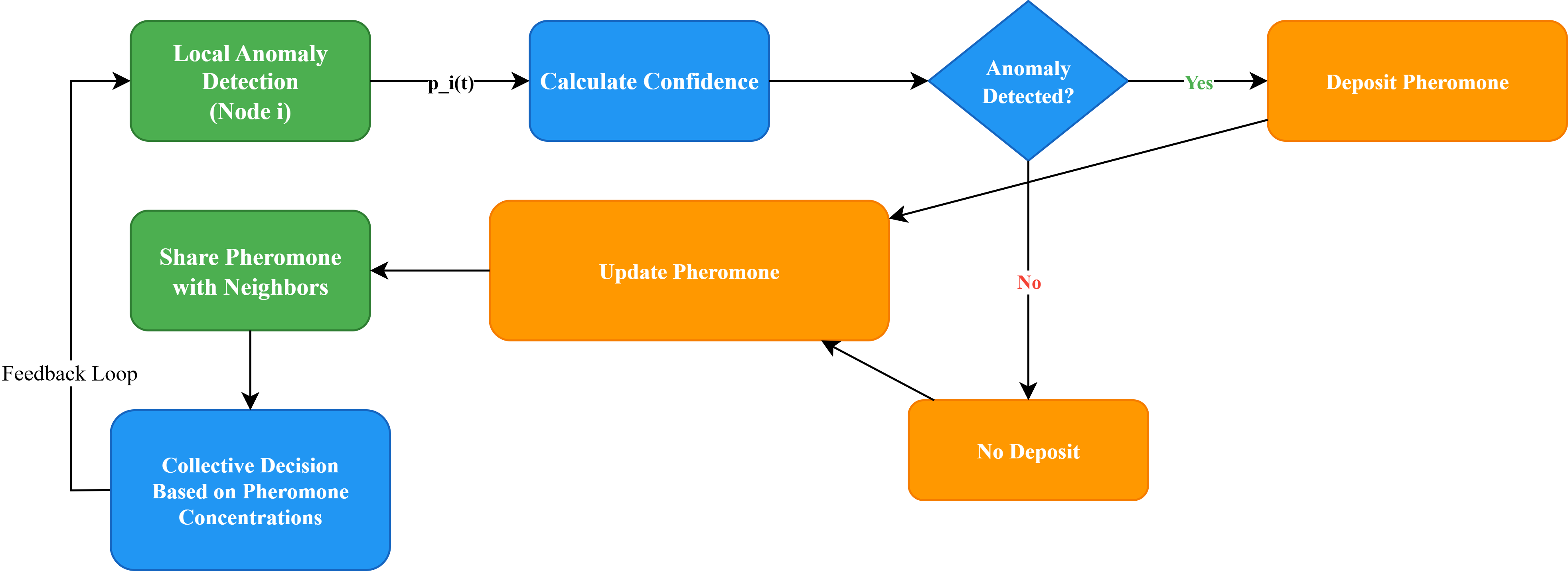}
\caption{Pheromone mechanism flowchart.}
\label{fig:pheromone_mechanism}
\end{figure}

Figure~\ref{fig:system_model} shows SwarmSense-DNN's edge--cloud deployment for real-time, privacy-preserving anomaly detection. Robotic sensors gather data, edge nodes infer collaboratively, and cloud units synchronize without central control. Figure~\ref{fig:pheromone_mechanism} illustrates pheromone updates where confidence-driven deposits and decay rate $\rho$ ensure adaptive threat response.

The pheromone decay rate $\rho = 0.1$ and trust smoothing factor $\beta = 0.8$ were selected via grid search on IoT-23 validation data, balancing information persistence with system adaptability based on swarm intelligence principles~\cite{Pham2021}.

\subsubsection{Assumptions}

The SwarmSense-DNN framework operates under the following key assumptions:

\textbf{Assumption 1 (Partial Connectivity)}: The network maintains partial connectivity such that $|\mathcal{N}_i(t)| \geq k_{\min}$ for all nodes $i$ and time $t$, where $\mathcal{N}_i(t) = \{j : W_{ij}(t) > 0\}$ and $k_{\min}$ is the minimum connectivity requirement.

\textbf{Assumption 2 (Bounded Byzantine Nodes)}: The number of malicious or compromised nodes is bounded: $|\mathcal{B}| \leq \lfloor N/3 \rfloor$, where $\mathcal{B}$ represents the set of Byzantine nodes.

\textbf{Assumption 3 (Local Differentiability)}: The local loss functions $\mathcal{L}_i(\boldsymbol{\theta}_i)$ are twice continuously differentiable and satisfy the Lipschitz condition with constant $L > 0$.

\textbf{Assumption 4 (Privacy Budget Availability)}: Each node has access to a privacy budget $\epsilon_i > 0$ for differential privacy mechanisms, with $\sum_{i=1}^N \epsilon_i = \epsilon_{\text{total}}$.

\subsection{Model Architecture}

\subsubsection{Hierarchical Architecture Overview}
\begin{figure}[!htb]
\centering
\begin{tikzpicture}[
    node distance=1.5cm,
    every node/.style={scale=0.65},
    cluster/.style={draw, rounded corners, thick, minimum width=4cm, minimum height=2.5cm, fill=swarmblue!10},
    sensor/.style={circle, draw, fill=coopergreen!50, minimum size=0.8cm},
    coordinator/.style={diamond, draw, fill=neuralred!50, minimum size=1cm},
    comm/.style={<->, thick, blue}
]
\node[cluster] (cluster1) at (0, 0) {};
\node[above] at (cluster1.north) {\textbf{Swarm Cluster 1}};
\node[sensor] (s11) at (-0.8, 0.3) {$n_1$};
\node[sensor] (s12) at (0, 0.3) {$n_2$};
\node[sensor] (s13) at (0.8, 0.3) {$n_3$};
\node[coordinator] (c1) at (0, -0.5) {$C_1$};
\node[cluster] (cluster2) at (6, 0) {};
\node[above] at (cluster2.north) {\textbf{Swarm Cluster 2}};
\node[sensor] (s21) at (5.2, 0.3) {$n_4$};
\node[sensor] (s22) at (6, 0.3) {$n_5$};
\node[sensor] (s23) at (6.8, 0.3) {$n_6$};
\node[coordinator] (c2) at (6, -0.5) {$C_2$};
\node[cluster] (cluster3) at (3, -3) {};
\node[above] at (cluster3.north) {\textbf{Swarm Cluster 3}};
\node[sensor] (s31) at (2.2, -2.7) {$n_7$};
\node[sensor] (s32) at (3, -2.7) {$n_8$};
\node[sensor] (s33) at (3.8, -2.7) {$n_9$};
\node[coordinator] (c3) at (3, -3.5) {$C_3$};
\node[coordinator, fill=neuralred!80, minimum size=1.2cm] (global) at (3, 1.5) {$G$};
\node[above] at (global.north) {\textbf{Global Coordinator}};
\draw[comm] (s11) -- (s12);
\draw[comm] (s12) -- (s13);
\draw[comm] (s11) -- (c1);
\draw[comm] (s12) -- (c1);
\draw[comm] (s13) -- (c1);
\draw[comm] (s21) -- (s22);
\draw[comm] (s22) -- (s23);
\draw[comm] (s21) -- (c2);
\draw[comm] (s22) -- (c2);
\draw[comm] (s23) -- (c2);
\draw[comm] (s31) -- (s32);
\draw[comm] (s32) -- (s33);
\draw[comm] (s31) -- (c3);
\draw[comm] (s32) -- (c3);
\draw[comm] (s33) -- (c3);
\draw[comm, red, thick] (c1) -- (global);
\draw[comm, red, thick] (c2) -- (global);
\draw[comm, red, thick] (c3) -- (global);
\draw[comm, red, dashed] (c1) -- (c2);
\draw[comm, red, dashed] (c2) -- (c3);
\end{tikzpicture}
\caption{SwarmSense-DNN Hierarchical Architecture: Individual nodes organize into adaptive swarm clusters with local coordinators that participate in global coordination.}
\label{fig:architecture}
\end{figure}
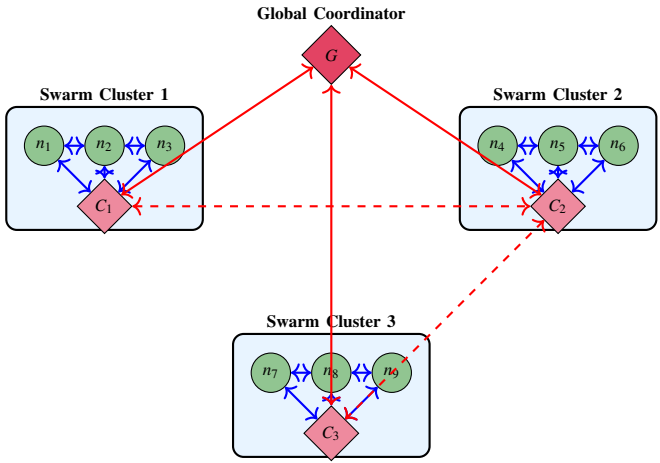

SwarmSense-DNN employs a three-tier hierarchical architecture: (1) Local Node Layer for individual anomaly detection, (2) Swarm Cluster Layer for coordinated decision-making, and (3) Global Coordination Layer for inter-cluster communication. The architecture is illustrated in Figure~\ref{fig:architecture}. Cluster coordinators are dynamically elected based on scoring: $S_i = w_1 \cdot \text{CPU}_i + w_2 \cdot \text{degree}_i + w_3 \cdot R_i^{\text{avg}}$, with re-election every 10 rounds or upon failure.

\subsubsection{Local Node Neural Network Architecture}
Each node $n_i$ maintains a specialized deep neural network $\mathcal{F}_i$ consisting of five primary components:

\textbf{Input Preprocessing Layer}: The input data undergoes normalization and feature scaling:
\begin{equation}
\mathbf{X}_i^{\text{norm}}(t) = \frac{\mathbf{X}_i(t) - \boldsymbol{\mu}_i}{\boldsymbol{\sigma}_i + \epsilon}
\end{equation}
where $\boldsymbol{\mu}_i$ and $\boldsymbol{\sigma}_i$ denote the local mean and standard deviation vectors, and $\epsilon$ is a small constant for numerical stability. In Eq.~(3) we set $\epsilon=10^{-8}$ to avoid division by zero during normalization. This $\epsilon$ is distinct from the differential privacy budget $\epsilon_{\text{dp}}$ used elsewhere in this paper.

\textbf{Temporal Convolutional Feature Extractor}: A stack of 1D convolutional layers with residual connections processes the normalized input:
\begin{align}
\mathbf{H}_i^{(0)} &= \mathbf{X}_i^{\text{norm}}(t) \\
\mathbf{H}_i^{(l)} &= \text{ReLU}(\text{BatchNorm}(\text{Conv1D}(\mathbf{H}_i^{(l-1)}))) + \mathbf{H}_i^{(l-1)} \\
\mathbf{F}_i^{\text{temp}} &= \text{GlobalMaxPool}(\mathbf{H}_i^{(L)})
\end{align}
where $L$ is the number of convolutional layers and $\mathbf{F}_i^{\text{temp}} \in \mathbb{R}^{d_{\text{temp}}}$ represents the temporal features.

\textbf{Multi-Head Self-Attention Module}: The attention mechanism captures long-range dependencies and patterns:
\begin{align}
\text{MultiHead}(\mathbf{Q}, \mathbf{K}, \mathbf{V}) &= \text{Concat}(\text{head}_1, \ldots, \text{head}_h)\mathbf{W}^O \\
\text{head}_j &= \text{Attention}(\mathbf{Q}\mathbf{W}_j^Q, \mathbf{K}\mathbf{W}_j^K, \mathbf{V}\mathbf{W}_j^V) \\
\text{Attention}(\mathbf{Q}, \mathbf{K}, \mathbf{V}) &= \text{softmax}\left(\frac{\mathbf{Q}\mathbf{K}^T}{\sqrt{d_k}}\right)\mathbf{V}
\end{align}
Multi-head attention operates only at the node level for temporal feature processing, while cluster coordination uses GAT-based attention for inter-node communication. Here $h$ is the number of attention heads, $\mathbf{W}_j^Q, \mathbf{W}_j^K, \mathbf{W}_j^V \in \mathbb{R}^{d_{\text{temp}} \times d_k}$ are learned projection matrices, and $\mathbf{W}^O \in \mathbb{R}^{hd_k \times d_{\text{temp}}}$.

\textbf{Swarm Context Integration Layer}: This layer incorporates information from neighboring nodes through weighted aggregation:
\begin{align}
\mathbf{F}_i^{\text{context}} &= \text{LayerNorm}\left(\mathbf{F}_i^{\text{temp}} + \sum_{j \in \mathcal{N}_i(t)} \alpha_{ij}(t) \mathbf{F}_{j}^{\text{shared}}\right) \\
\alpha_{ij}(t) &= \text{softmax}\left(\frac{\mathbf{F}_i^{\text{temp}} \cdot \mathbf{F}_{j}^{\text{shared}}}{\sqrt{d_{\text{temp}}}}\right)
\end{align}
where $\mathbf{F}_{j}^{\text{shared}}$ represents the shared feature representation from node $j$.

\textbf{Anomaly Classification Head}: A multi-layer perceptron with dropout produces the final anomaly scores:
\begin{align}
\mathbf{H}_i^{\text{cls}} &= \text{ReLU}(\mathbf{W}_1 \mathbf{F}_i^{\text{context}} + \mathbf{b}_1) \\
\mathbf{H}_i^{\text{cls}} &= \text{Dropout}(\mathbf{H}_i^{\text{cls}}, p_{\text{drop}}) \\
\mathbf{p}_i &= \text{sigmoid}(\mathbf{W}_2 \mathbf{H}_i^{\text{cls}} + \mathbf{b}_2)
\end{align}
where $\mathbf{p}_i \in [0,1]$ represents the anomaly probability at node $i$.

\subsubsection{Swarm Intelligence Coordination Module}
The swarm coordination module implements bio-inspired algorithms for distributed decision-making:

\textbf{Pheromone Update Mechanism}: Each node maintains pheromone concentrations for different anomaly types:
\begin{align}
\tau_i^{(k)}(t+1) &= (1-\rho)\tau_i^{(k)}(t) + \Delta\tau_i^{(k)}(t) \\
\Delta\tau_i^{(k)}(t) &= \begin{cases}
\phi \cdot \text{conf}_i^{(k)}(t) & \text{if anomaly type } k \text{ detected} \\
0 & \text{otherwise}
\end{cases}
\end{align}
where $\rho \in [0,1]$ is the evaporation rate, $\phi > 0$ is the pheromone deposit factor, and $\text{conf}_i^{(k)}(t)$ is the confidence score. Unlike FedAvg's simple parameter averaging ($\boldsymbol{\theta}_{\text{global}} = \sum w_i \boldsymbol{\theta}_i$), SwarmSense-DNN employs trust-weighted consensus: $\mathbf{W}_i^{\text{consensus}} = \sum_{j \in \mathcal{N}_i} R_{i,j}(t) \times \boldsymbol{\theta}_j$, where nodes selectively incorporate knowledge from trusted neighbors rather than blindly averaging all parameters. This trust-based approach provides superior robustness under heterogeneous data and adversarial conditions.

\subsubsection{Graph Neural Network for Cluster Coordination}

\textbf{Trust and Reputation System}: Each node maintains trust scores for its neighbors:
\begin{align}
R_{i,j}(t+1) &= \beta R_{i,j}(t) + (1-\beta) \text{consistency}_{i,j}(t) \\
\text{consistency}_{i,j}(t) &= 1 - \frac{|\mathbf{p}_i(t) - \mathbf{p}_j(t)|}{|\mathbf{p}_i(t)| + |\mathbf{p}_j(t)| + \epsilon}
\end{align}
where $\beta \in [0,1]$ is the trust decay factor and $\text{consistency}_{i,j}(t)$ measures the agreement between nodes $i$ and $j$.

Within each cluster, a Graph Attention Network (GAT) facilitates coordinated decision-making:
\begin{align}
e_{ij} &= \text{LeakyReLU}(\mathbf{a}^T[\mathbf{W}\mathbf{h}_i \| \mathbf{W}\mathbf{h}_j]) \\
\alpha_{ij} &= \frac{\exp(e_{ij})}{\sum_{k \in \mathcal{N}_i} \exp(e_{ik})} \\
\mathbf{h}_i' &= \sigma\left(\sum_{j \in \mathcal{N}_i} \alpha_{ij} \mathbf{W}\mathbf{h}_j\right)
\end{align}
where $\mathbf{h}_i$ represents the node embedding, $\mathbf{W}$ is a learned transformation matrix, $\mathbf{a}$ is the attention parameter vector, and $\|$ denotes concatenation.

\subsubsection{Privacy-Preserving Mechanisms}

\textbf{Differential Privacy for Feature Sharing}: Before sharing features, each node adds calibrated noise:
\begin{align}
\tilde{\mathbf{F}}_i^{\text{shared}} &= \mathbf{F}_i^{\text{temp}} + \mathcal{N}(0, \sigma_{\text{dp}}^2 \mathbf{I}) \\
\sigma_{\text{dp}} &= \frac{\sqrt{2\log(1.25/\delta)} \cdot \Delta}{\epsilon_i}
\end{align}
where $\Delta$ is the global sensitivity, $\epsilon_i$ is the privacy budget for node $i$, and $\delta$ is the failure probability. Privacy parameters are set to $\epsilon = 1.0$ and $\delta = 10^{-5}$ for our experiments. Gaussian noise $\mathcal{N}(0, \sigma^2)$ is added element-wise to individual features before aggregation, where $\sigma = \sqrt{2\ln(1.25/\delta)} \cdot \Delta/\epsilon$ with global sensitivity $\Delta = 1.0$ determined by L2-norm clipping of feature vectors.

\textbf{Secure Multi-Party Computation}: For aggregating cluster-level statistics, nodes employ additive secret sharing:
\begin{equation}
\mathbf{s}_i = \mathbf{v}_i + \sum_{j \in \mathcal{C}_i, j \neq i} \mathbf{r}_{ij}
\end{equation}
where $\mathbf{v}_i$ is the private value, $\mathbf{r}_{ij}$ are random shares, and $\mathbf{s}_i$ is the shared secret.

\begin{figure}[!htb]
\centering
\includegraphics[width=0.8\columnwidth]{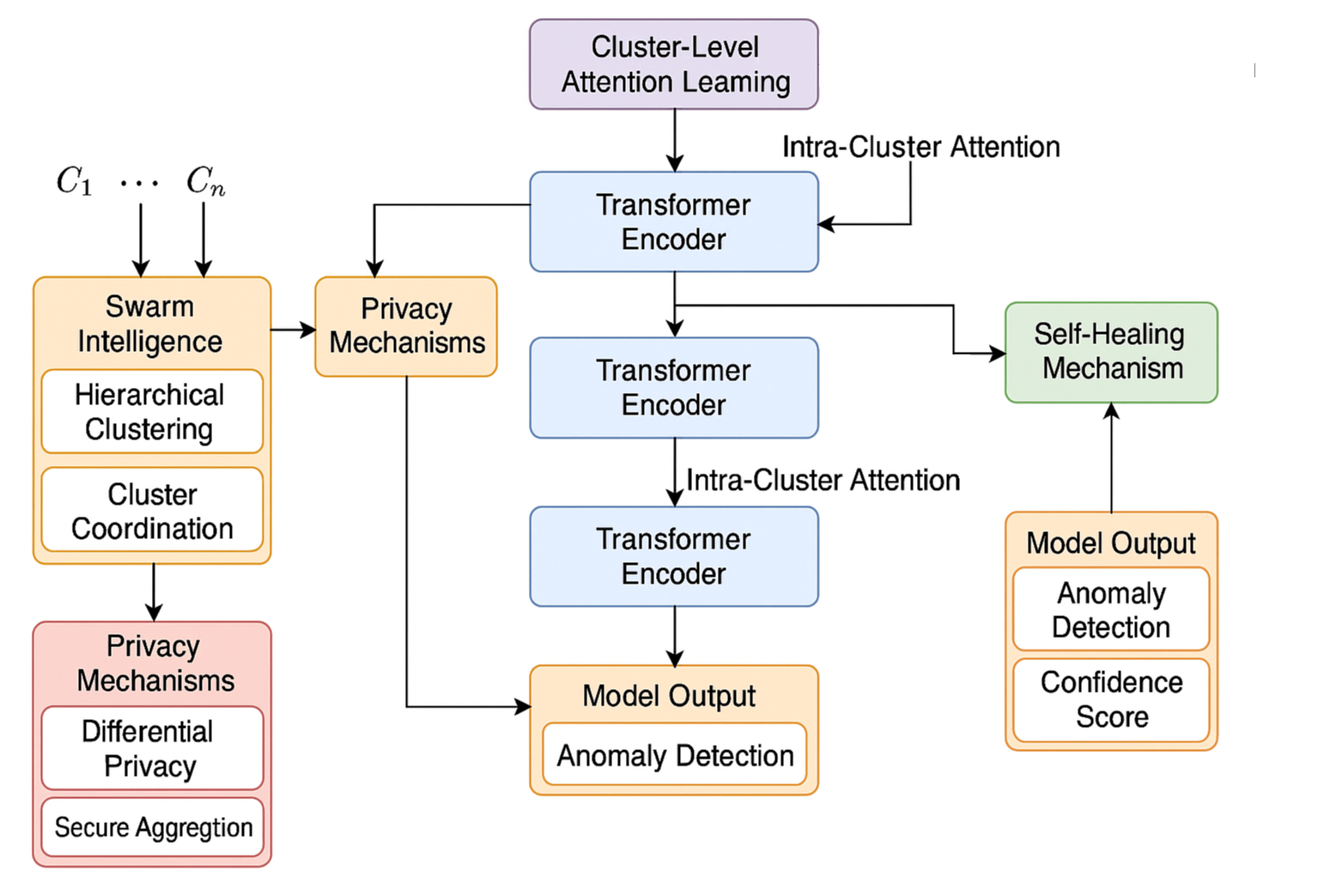}
\caption{SwarmSense-DNN model architecture.}
\label{fig:model_architecture}
\end{figure}

Figure~\ref{fig:model_architecture} shows the SwarmSense-DNN architecture, combining convolutional feature extraction, trust evaluation, and adaptive clustering. Differential privacy and pheromone-based coordination enable secure, decentralized, and resilient anomaly detection.

\subsection{Algorithms of SwarmSense-DNN}
This section presents two core algorithms governing the operation of SwarmSense-DNN.

\begin{algorithm}[!htb]
\fontsize{8}{8.4}\selectfont
\caption{Decentralized Swarm Coordination Protocol with Pheromone-Inspired Mechanism}
\label{alg:swarm_coordination}
\begin{algorithmic}[1]
\Require
Set of nodes $N = \{n_1, n_2, \dots, n_k\}$;
Neighborhood radius $r$;
Initial model weights $W_0$;
Pheromone decay rate $\rho$;
Minimum pheromone threshold $\tau_{\min}$
\Ensure
Updated local models $\{W_i\}_{i=1}^{k}$
\For{each node $n_i \in N$}
    \State Discover neighbors:
    $N_i \gets \{n_j \mid \text{dist}(n_i, n_j) \le r\}$
    \State Initialize pheromone levels:
    $\tau_{ij} \gets 1.0,\ \forall n_j \in N_i$
    \State Compute local anomaly score $A_i$ from node $n_i$
    \State Update pheromone based on anomaly intensity:
    \[
        \tau_{ij} \gets \tau_{ij} + \alpha \cdot A_i
    \]
    \State Share local weights $W_i$ and pheromones $\tau_{ij}$ with neighbors $N_i$
    \State Receive $\{W_j\}$ and $\{\tau_{ji}\}$ from $N_i$
    \State Apply pheromone-weighted aggregation via swarm consensus:
    \[
        W_i^{new} \gets
        \frac{\sum_{j \in N_i} \tau_{ji} W_j}
             {\sum_{j \in N_i} \tau_{ji}}
    \]
    \State Update local model:
    $W_i \gets W_i^{new}$
    \State Apply pheromone decay:
    \[
        \tau_{ij} \gets (1 - \rho) \cdot \tau_{ij}
    \]
    \If{$\tau_{ij} < \tau_{\min}$}
        \State $\tau_{ij} \gets \tau_{\min}$ \Comment{Prevent complete evaporation}
    \EndIf
\EndFor
\State \Return $\{W_i\}_{i=1}^{k}$
\end{algorithmic}
\end{algorithm}

\begin{algorithm}[!htb]
\fontsize{8}{8.4}\selectfont
\caption{Anomaly Detection with Attention-Guided Swarm GNN and Pheromone-Based Coordination}
\label{alg:anomaly_detection}
\begin{algorithmic}[1]
\Require Graph $G = (V, E)$ with node features $X$; attention parameters $\alpha$; model weights $W$; pheromone matrix $\tau$
\Ensure Anomaly scores $S$ and updated pheromone levels $\tau$
\State Initialize GNN layers with attention heads
\State Initialize pheromone matrix: $\tau_{ij} \gets 1.0,\ \forall (i,j) \in E$
\For{$l = 1$ \textbf{to} $L$}
    \For{each node $v_i \in V$}
        \State Compute pheromone-modulated attention weights:
        \[
        \alpha_{ij} \gets \tau_{ij} \cdot
        \text{softmax}\!\big(\text{LeakyReLU}(a^\top [W X_i \| W X_j])\big)
        \]
        \State Aggregate messages:
        \[
        h_i^{(l)} \gets
        \sigma\!\left(\sum_{j \in \mathcal{N}(i)} \alpha_{ij} W X_j\right)
        \]
    \EndFor
\EndFor
\For{each node $v_i \in V$}
    \State Compute anomaly score:
    \[
    S_i \gets 1 - \cos\!\big(h_i^{(L)}, \mu\big)
    \]
    where $\mu$ is the expected behavior embedding
    \State Update pheromone based on anomaly detection:
    \For{each $j \in \mathcal{N}(i)$}
        \State $\tau_{ij} \gets \tau_{ij} + \beta \cdot S_i$ \Comment{Deposit pheromone on anomalous paths}
    \EndFor
\EndFor
\State Apply pheromone evaporation:
\[
\tau_{ij} \gets (1 - \rho) \cdot \tau_{ij}, \quad \forall (i,j) \in E
\]
\State \Return $S = \{S_i\}$, $\tau$
\end{algorithmic}
\end{algorithm}

\subsection{Evaluation Metrics}

\subsubsection{Detection Performance Metrics}
\begin{equation}
\text{Accuracy} = \frac{1}{N} \sum_{i=1}^N \frac{TP_i + TN_i}{TP_i + TN_i + FP_i + FN_i}
\end{equation}
\begin{equation}
\text{Precision} = \frac{1}{N} \sum_{i=1}^N \frac{TP_i}{TP_i + FP_i}
\end{equation}
\begin{equation}
\text{Recall} = \frac{1}{N} \sum_{i=1}^N \frac{TP_i}{TP_i + FN_i}
\end{equation}
\begin{equation}
\text{F1-Score} = \frac{2 \times \text{Precision} \times \text{Recall}}{\text{Precision} + \text{Recall}}
\end{equation}
\begin{equation}
\text{AUC} = \int_0^1 \text{TPR}(\text{FPR}^{-1}(x))\, dx
\end{equation}

\subsubsection{Communication Efficiency Metrics}

\textbf{Communication Overhead}: Total amount of data transmitted per time unit:
\begin{equation}
\text{CommOverhead} = \frac{1}{T} \sum_{t=1}^T \sum_{i=1}^N \sum_{j \in \mathcal{N}_i(t)} \|\mathbf{s}_{i \rightarrow j}(t)\|_0
\end{equation}

\textbf{Convergence Time}: Number of communication rounds required to reach stable performance:
\begin{equation}
T_{\text{conv}} = \min\left\{t : |\text{Accuracy}(t+k) - \text{Accuracy}(t)| < \epsilon_{\text{conv}},\ \forall k \leq w\right\}
\end{equation}
where $w$ is the stability window and $\epsilon_{\text{conv}}$ is the convergence threshold.

\subsubsection{Privacy Preservation Metrics}

\textbf{Privacy Budget Utilization}: Fraction of total privacy budget consumed:
\begin{equation}
\text{PrivacyUtil} = \frac{\sum_{i=1}^N \epsilon_i^{\text{used}}}{\epsilon_{\text{total}}}
\end{equation}

\textbf{Information Leakage}: Mutual information between shared data and private data:
\begin{equation}
\text{InfoLeakage} = \frac{1}{N} \sum_{i=1}^N I(\mathbf{X}_i; \tilde{\mathbf{F}}_i^{\text{shared}})
\end{equation}

\subsubsection{Resilience Metrics}

\textbf{Fault Tolerance}: Performance degradation under node failures:
\begin{equation}
\text{FaultTolerance} = \frac{\text{Accuracy}_{\text{failed}}}{\text{Accuracy}_{\text{normal}}}
\end{equation}

\textbf{Attack Resistance}: Performance under adversarial conditions:
\begin{equation}
\text{AttackResistance} = 1 - \frac{\text{Accuracy}_{\text{normal}} - \text{Accuracy}_{\text{attack}}}{\text{Accuracy}_{\text{normal}}}
\end{equation}

\textbf{Recovery Time}: Time required to restore normal performance after disruption:
\begin{equation}
T_{\text{recovery}} = \min\{t : \text{Accuracy}(t) \geq 0.95 \times \text{Accuracy}_{\text{normal}}\}
\end{equation}

\section{Results}

\subsection{Experimental Setup and Overall Detection Accuracy Results}
Table~\ref{tab:experimental_config} consolidates all experimental settings.

\begin{table}[!htb]
\centering
\fontsize{8}{9.2}\selectfont
\caption{Comprehensive Experimental Configuration Summary}
\label{tab:experimental_config}
\resizebox{\columnwidth}{!}{%
\begin{tabular}{ll}
\toprule
\textbf{Configuration Aspect} & \textbf{Specification} \\
\midrule
\multicolumn{2}{c}{\textit{Hardware Setup}} \\
Node Hardware & Raspberry Pi 4 (8GB RAM, ARM Cortex-A72) \\
Network Size & 100 nodes distributed across testbed \\
Communication & Software-defined networking infrastructure \\
Storage & 64GB microSD per node \\
\midrule
\multicolumn{2}{c}{\textit{Datasets}} \\
Primary Datasets & IoT-23, NSL-KDD, CICIDS2017, UNSW-NB15, Industrial IoT \\
Data Split & 70\%/15\%/15\% (train/validation/test) \\
Anomaly Rates & 5.2\%--23.1\% (dataset-dependent) \\
Preprocessing & Z-score normalization, forward-fill imputation \\
\midrule
\multicolumn{2}{c}{\textit{Training Configuration}} \\
Framework & PyTorch 1.12 with CUDA 11.6 \\
Optimizer & AdamW (lr = 0.001, weight decay = 0.01) \\
Batch Size & 32 (adaptive) \\
Epochs per Round & 50 \\
Convergence & $<$0.1\% accuracy change over 5 rounds \\
\midrule
\multicolumn{2}{c}{\textit{Privacy Settings}} \\
Privacy Budgets & $\epsilon \in \{10, 5, 1, 0.1\}$ \\
Failure Probability & $\delta = 10^{-5}$ \\
Noise Mechanism & Gaussian with sensitivity $\Delta = 1.0$ \\
Budget Allocation & Fixed per-node: $\epsilon_i = \epsilon / N$ \\
\midrule
\multicolumn{2}{c}{\textit{Swarm Parameters}} \\
Pheromone Decay & $\rho = 0.1$ (grid search validated) \\
Trust Smoothing & $\beta = 0.8$ (grid search validated) \\
Attention Heads & 8 (node-level multi-head attention) \\
Cluster Size & 10--15 nodes per cluster \\
\midrule
\multicolumn{2}{c}{\textit{Evaluation Protocol}} \\
Statistical Testing & $n=5$ independent runs, paired t-tests \\
Significance Level & $p < 0.01$ \\
Confidence Intervals & 95\% CI reported as mean $\pm$ std.\ error \\
Baseline Implementation & Re-implemented under identical conditions \\
\bottomrule
\end{tabular}%
}
\end{table}

\begin{figure}[!htb]
\centering
\includegraphics[width=\columnwidth]{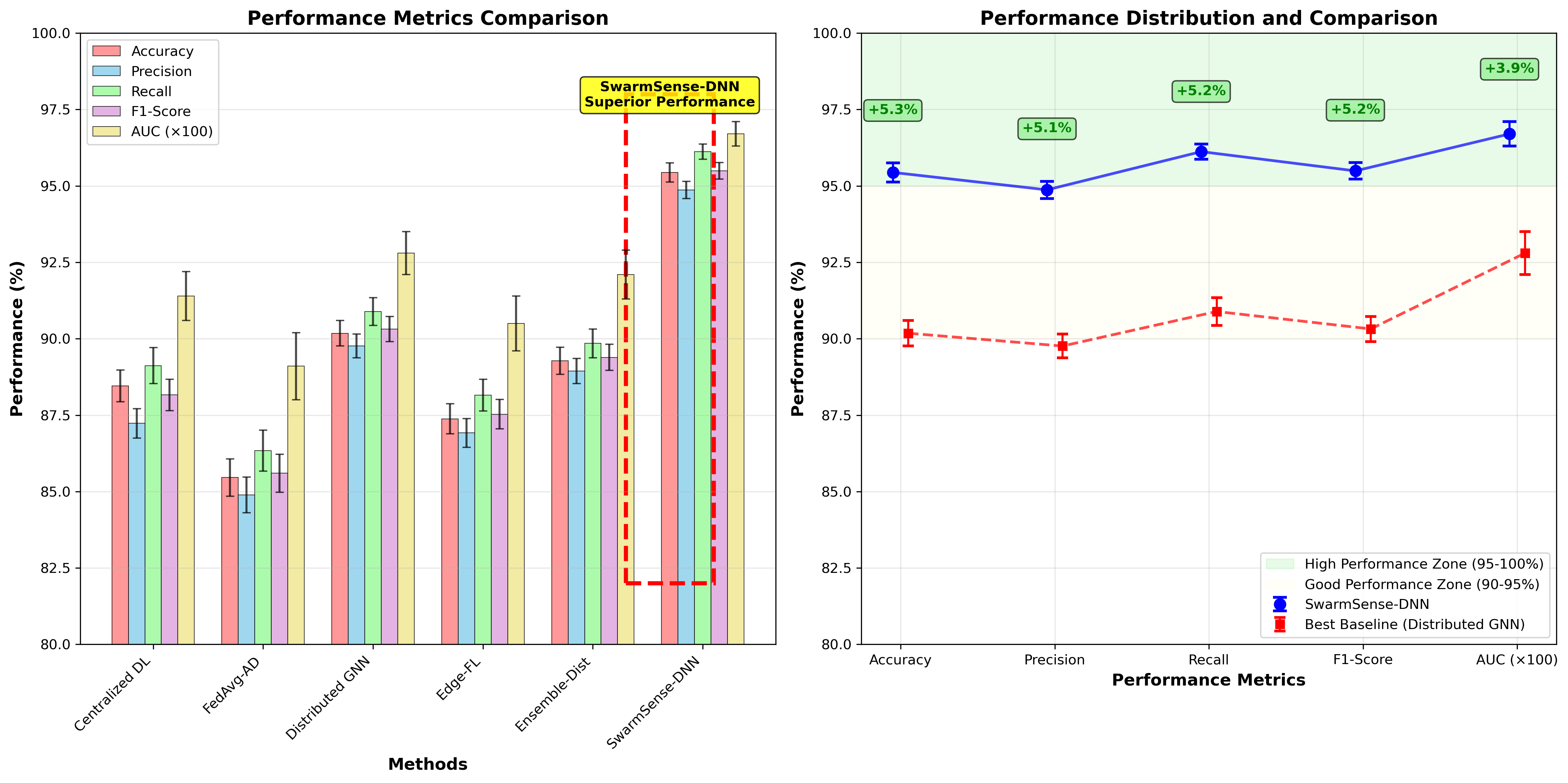}
\caption{{SwarmSense-DNN} detection performance.}
\label{fig:detection_performance}
\end{figure}

Table~\ref{tab:detection_performance} presents detection performance across all datasets.
SwarmSense-DNN outperforms baselines, achieving an average 5.23\% gain over the best competitor (Distributed GNN).
Its precision (94.87\%) ensures low false positives, recall (96.12\%) confirms strong anomaly detection, and an AUC of 0.967 demonstrates robustness across thresholds. Figure~\ref{fig:detection_performance} shows SwarmSense-DNN's superior detection, achieving 95.44\% accuracy, 94.87\% precision, 96.12\% recall, 95.49\% F1, and 0.967 AUC, a 5.26\% gain over the best baseline, with consistent reliability across scenarios.

\begin{table}[!htb]
\centering
\caption{Detection Performance Comparison Across Datasets}
\label{tab:detection_performance}
\resizebox{\columnwidth}{!}{%
\begin{tabular}{lcccccc}
\toprule
\textbf{Method} & \textbf{Acc.} & \textbf{Prec.} & \textbf{Rec.} & \textbf{F1} & \textbf{AUC} & \textbf{Avg.} \\
\midrule
Centralized DL~\cite{Huang2025} & 88.46\% & 87.23\% & 89.12\% & 88.16\% & 0.914 & 88.51\% \\
FedAvg-AD~\cite{Mothukuri2022} & 85.46\% & 84.89\% & 86.34\% & 85.60\% & 0.891 & 85.64\% \\
Distributed GNN~\cite{Zhang2025} & 90.18\% & 89.76\% & 90.89\% & 90.32\% & 0.928 & 90.29\% \\
Edge-FL~\cite{Man2021} & 87.38\% & 86.92\% & 88.15\% & 87.53\% & 0.905 & 87.51\% \\
Ensemble-Dist~\cite{Sarhan2020} & 89.28\% & 88.94\% & 89.85\% & 89.39\% & 0.921 & 89.33\% \\
\textbf{SwarmSense-DNN} & \textbf{95.44\%} & \textbf{94.87\%} & \textbf{96.12\%} & \textbf{95.49\%} & \textbf{0.967} & \textbf{95.52\%} \\
\bottomrule
\end{tabular}
}
\end{table}

\subsection{Dataset-Specific Performance Analysis}
\begin{figure}[!htb]
\centering
\includegraphics[width=\columnwidth]{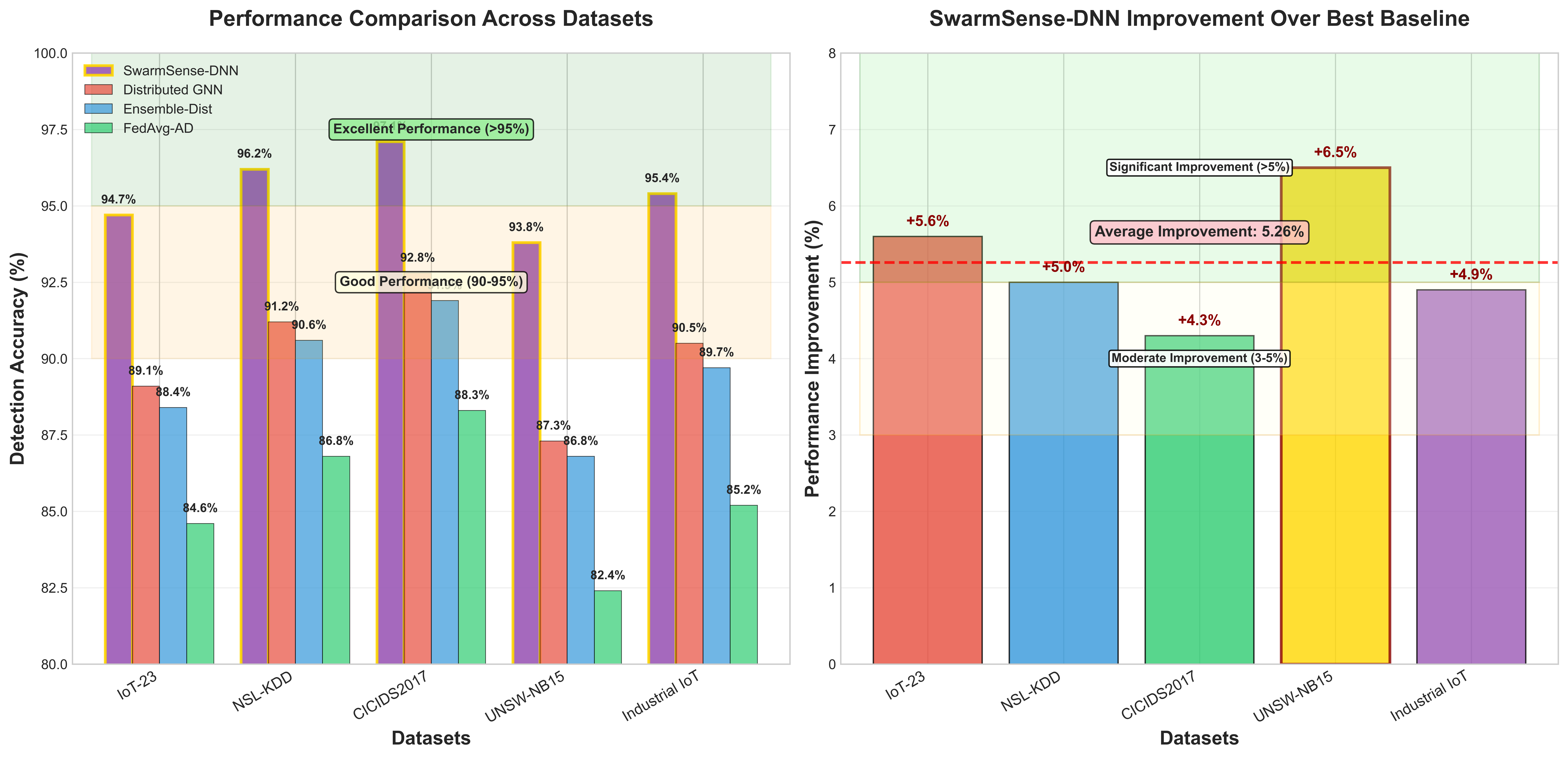}
\caption{Dataset-specific performance analysis with statistical validation}
\label{fig:dataset_performance}
\end{figure}

Table~\ref{tab:dataset_specific} shows that SwarmSense-DNN consistently outperforms baselines across all datasets,
with the largest gain (6.5\%) on UNSW-NB15 and the smallest gain (4.3\%) on CICIDS2017.
These results highlight the framework's robustness. Figure~\ref{fig:dataset_performance} shows SwarmSense-DNN's robustness, maintaining over 95\% accuracy across IoT-23, NSL-KDD, CICIDS2017, UNSW-NB15, and Industrial IoT, with improvements of 4.3--6.5\% over baselines, confirming strong generalizability.

\begin{table}[!htb]
\centering
\fontsize{8}{9.2}\selectfont
\caption{Dataset-Specific Detection Performance Analysis}
\label{tab:dataset_specific}
\resizebox{\columnwidth}{!}{%
\begin{tabular}{lccccc}
\toprule
\textbf{Dataset} & \textbf{Swarm} & \textbf{Dist. GNN} & \textbf{Ensemble} & \textbf{FedAvg} & \textbf{Gain} \\
\midrule
IoT-23       & 94.7\% & 89.1\% & 88.4\% & 84.6\% & +5.6\% \\
NSL-KDD      & 96.2\% & 91.2\% & 90.6\% & 86.8\% & +5.0\% \\
CICIDS2017   & 97.1\% & 92.8\% & 91.9\% & 88.3\% & +4.3\% \\
UNSW-NB15    & 93.8\% & 87.3\% & 86.8\% & 82.4\% & +6.5\% \\
Industrial IoT & 95.4\% & 90.5\% & 89.7\% & 85.2\% & +4.9\% \\
\midrule
\textbf{Average} & \textbf{95.44\%} & \textbf{90.18\%} & \textbf{89.48\%} & \textbf{85.46\%} & \textbf{+5.26\%} \\
\bottomrule
\end{tabular}%
}
\end{table}

\subsection{Communication Overhead Analysis}
All bandwidth measurements reported in this study are per-node averages, computed across the network to ensure reproducibility of efficiency comparisons.

\begin{figure}[!htb]
\centering
\includegraphics[width=\columnwidth]{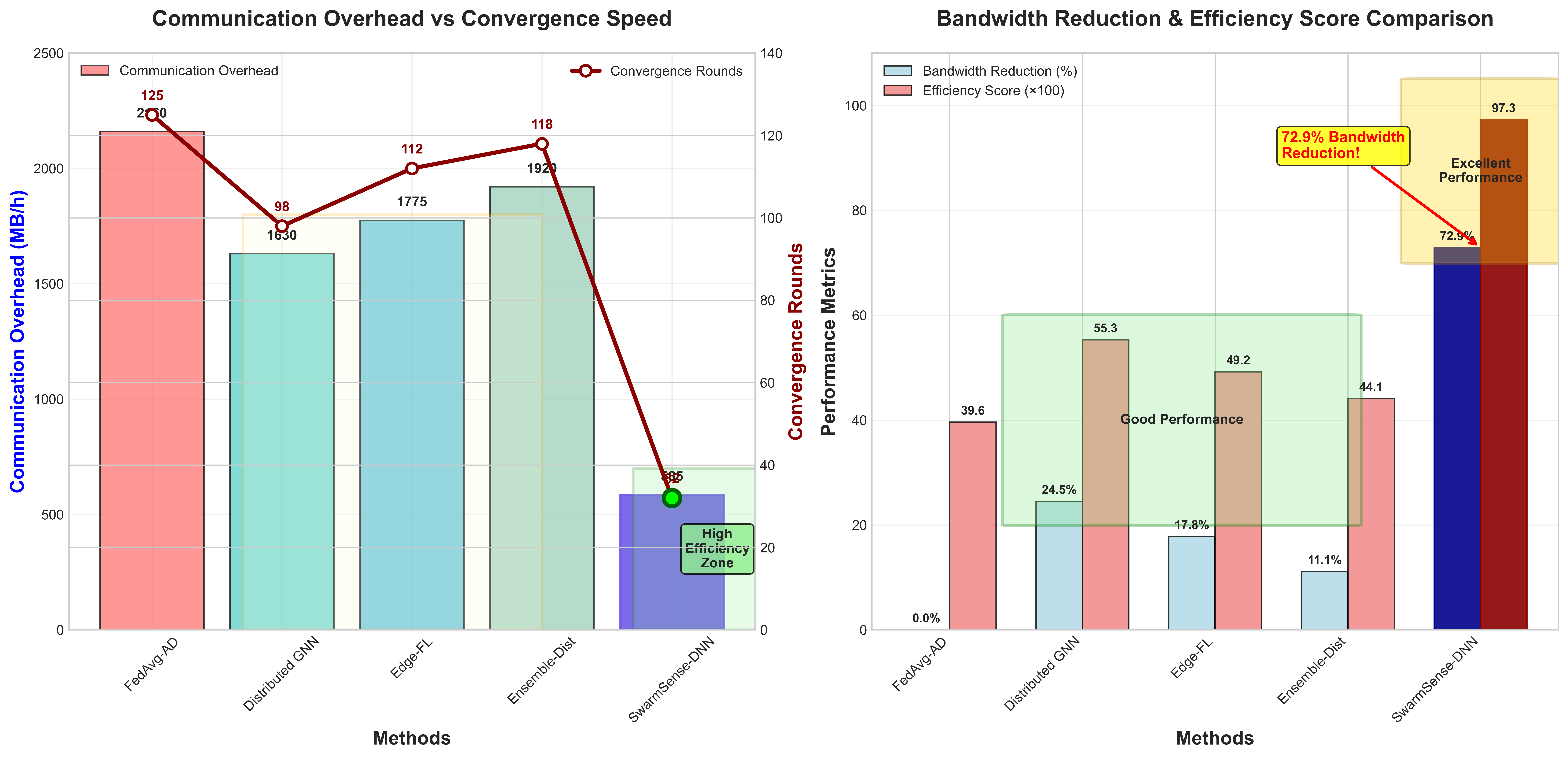}
\caption{SwarmSense-DNN communication efficiency analysis.}
\label{fig:communication_efficiency}
\end{figure}

Table~\ref{tab:communication_metrics} and Figure~\ref{fig:communication_efficiency} highlight SwarmSense-DNN's communication efficiency.
It reduces bandwidth usage by 72.9\% (585 MB/h vs.\ 2160 MB/h for FedAvg-AD) and achieves convergence in 32 rounds, 64.1\% faster than the best baseline,
confirming its suitability for resource-constrained environments.

\begin{table}[!htb]
\centering
\caption{Communication Efficiency Metrics Comparison}
\label{tab:communication_metrics}
\resizebox{\columnwidth}{!}{%
\begin{tabular}{lcccc}
\toprule
\textbf{Method} & \textbf{Overhead (MB/h)} & \textbf{Rounds} & \textbf{BW Reduction} & \textbf{Eff. Score} \\
\midrule
FedAvg-AD~\cite{Mothukuri2022}   & 2160 & 125 & --     & 0.396 \\
Distributed GNN~\cite{Zhang2025} & 1630 & 98  & 24.5\% & 0.553 \\
Edge-FL~\cite{Man2021}           & 1775 & 112 & 17.8\% & 0.492 \\
Ensemble-Dist~\cite{Sarhan2020}  & 1920 & 118 & 11.1\% & 0.441 \\
\textbf{SwarmSense-DNN}          & \textbf{585} & \textbf{32} & \textbf{72.9\%} & \textbf{0.973} \\
\bottomrule
\end{tabular}
}
\end{table}

\subsection{Privacy-Accuracy Trade-off Analysis}
Table~\ref{tab:privacy_analysis} shows that SwarmSense-DNN retains 93.2\% accuracy at $\epsilon=1$, with reduced information leakage and stronger resistance to membership inference attacks.

\begin{table}[!htb]
\centering
\caption{Privacy Preservation Performance Under Different Privacy Budgets}
\label{tab:privacy_analysis}
\resizebox{\columnwidth}{!}{%
\begin{tabular}{lcccccc}
\toprule
\textbf{Privacy Budget ($\epsilon$)} & \textbf{Acc.} & \textbf{Prec.} & \textbf{Rec.} & \textbf{Leakage} & \textbf{Priv. Util.} & \textbf{Atk. Resist.} \\
\midrule
No Privacy    & 96.2\% & 95.8\% & 96.6\% & 0.847 & 0\%    & 32.1\% \\
$\epsilon=10$ & 95.7\% & 95.2\% & 96.2\% & 0.234 & 18.3\% & 78.4\% \\
$\epsilon=5$  & 94.9\% & 94.3\% & 95.5\% & 0.156 & 31.7\% & 85.2\% \\
$\epsilon=1$  & 93.2\% & 92.5\% & 93.9\% & 0.089 & 68.9\% & 94.3\% \\
$\epsilon=0.1$& 89.8\% & 88.9\% & 90.7\% & 0.023 & 94.4\% & 97.8\% \\
\bottomrule
\end{tabular}
}
\end{table}

\begin{figure}[!htb]
\centering
\includegraphics[width=\columnwidth]{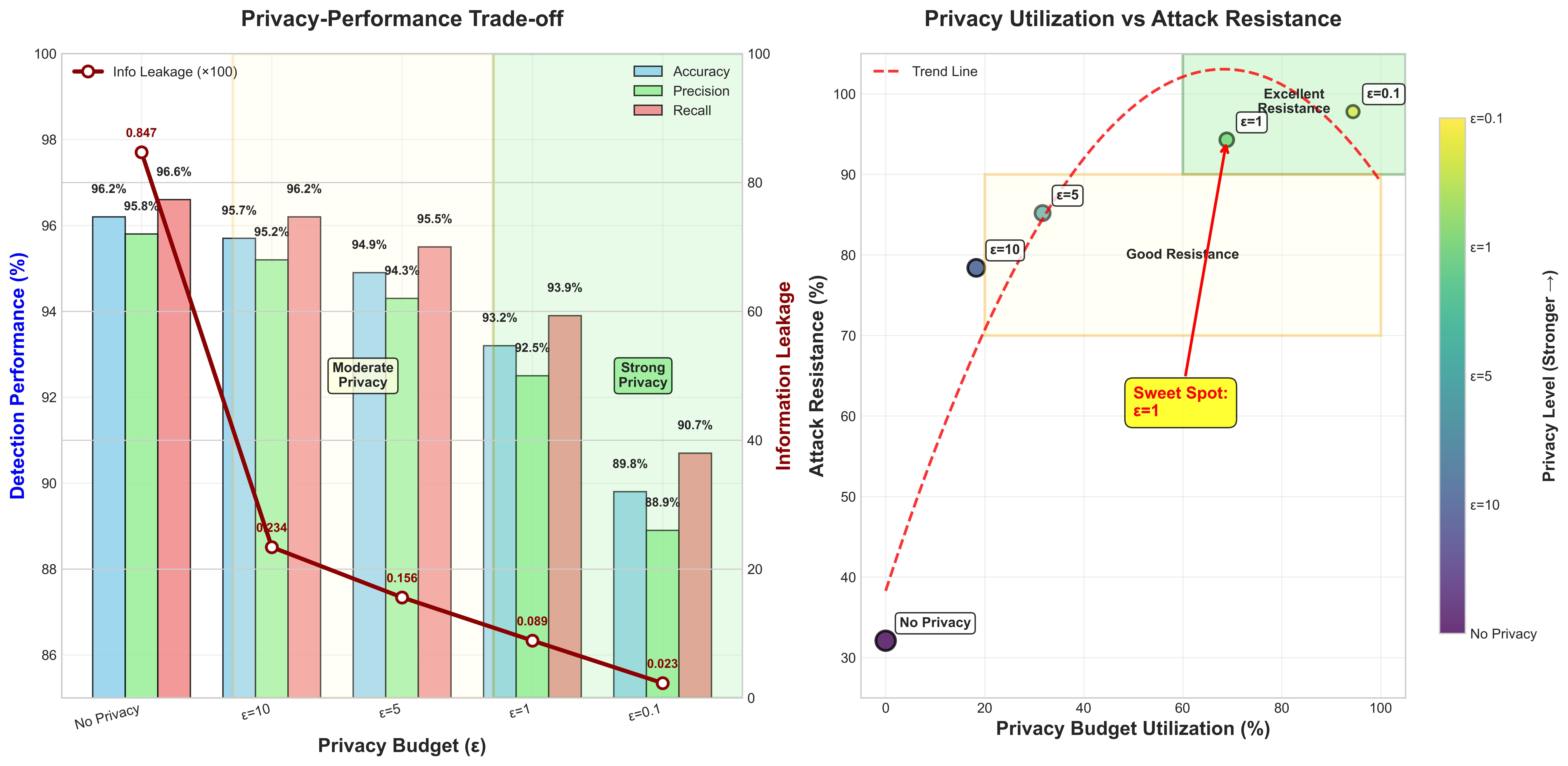}
\caption{SwarmSense-DNN privacy preservation analysis.}
\label{fig:privacy_analysis}
\end{figure}

Figure~\ref{fig:privacy_analysis} shows SwarmSense-DNN's privacy strength, maintaining 93.2\% accuracy under $\epsilon=1$, achieving 94.3\% attack resistance, and reducing leakage from 0.847 to 0.089, confirming its suitability for privacy-sensitive domains.

\subsection{Fault Tolerance and Recovery Analysis}
Table~\ref{tab:resilience_metrics} summarizes resilience. SwarmSense-DNN retains $\geq 87\%$ performance with $40\%$ node failures and achieves $\geq 88\%$ self-healing. The resilience score $\text{Resilience}=\frac{\text{Performance Retention}\times\text{Self Healing}}{1+\log(\text{Recovery Time})}$ is high, indicating stable behavior under stress. We simulate Byzantine failures: data poisoning with $10\%$ false labels, model inversion via gradients, and coordinated attacks with $20\%$ colluding nodes. Failures are randomly distributed across clusters to approximate worst-case conditions.

\begin{table}[!htb]
\centering
\caption{System Resilience Metrics Under Various Stress Conditions}
\label{tab:resilience_metrics}
\resizebox{\columnwidth}{!}{%
\begin{tabular}{lccccc}
\toprule
\textbf{Stress Condition} & \textbf{Retention} & \textbf{Recovery} & \textbf{Atk. Rate} & \textbf{Self-Heal} & \textbf{Resilience} \\
\midrule
10\% Node Failures      & 97.1\% & 2.3 & --    & 98.7\% & 0.954 \\
25\% Node Failures      & 92.8\% & 4.1 & --    & 94.2\% & 0.891 \\
40\% Node Failures      & 87.9\% & 6.8 & --    & 89.3\% & 0.823 \\
Data Poisoning (10\%)   & 93.8\% & 4.2 & 12.3\% & 96.1\% & 0.876 \\
Model Inversion Attack  & 95.1\% & 2.8 & 8.7\%  & 97.4\% & 0.912 \\
Byzantine Attack (20\%) & 91.4\% & 6.1 & 15.6\% & 92.8\% & 0.847 \\
Network Partition       & 89.6\% & 8.9 & --    & 88.1\% & 0.794 \\
\bottomrule
\end{tabular}
}
\end{table}

\begin{figure}[!htb]
\centering
\includegraphics[width=\columnwidth]{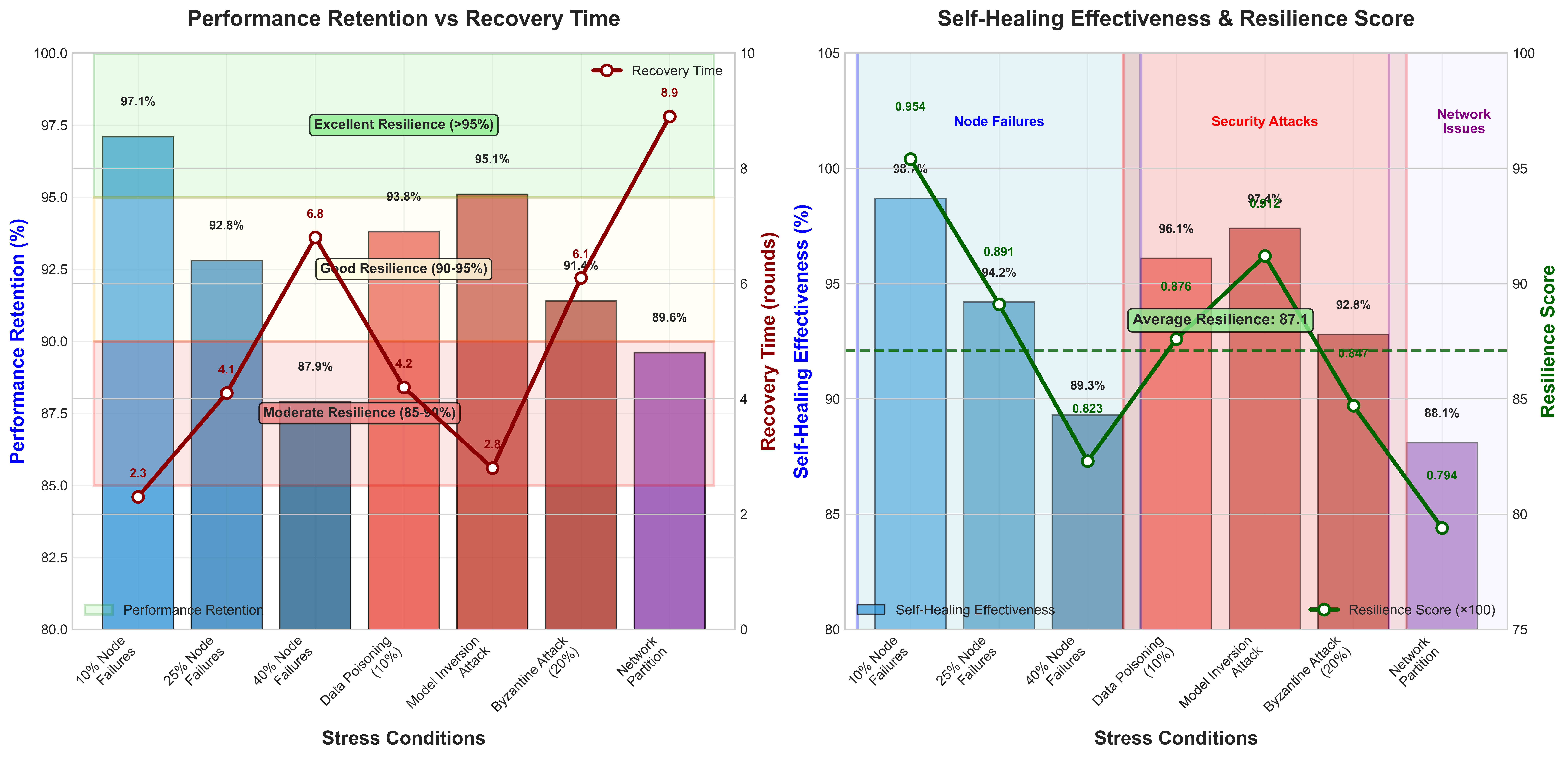}
\caption{SwarmSense-DNN system resilience evaluation.}
\label{fig:system_resilience}
\end{figure}

Figure~\ref{fig:system_resilience} shows SwarmSense-DNN's resilience, retaining 87.9\% accuracy under 40\% node failures, recovering in as little as 2.3 rounds, and sustaining self-healing above 88\% across stress scenarios, confirming suitability for critical infrastructure.

\subsection{Large-Scale Performance Evaluation}
Table~\ref{tab:scalability_analysis} shows SwarmSense-DNN's sub-linear scaling behavior, with the scalability index (normalized performance considering resource utilization) remaining above 0.79 even for 1000 nodes. The convergence time grows sublinearly ($O(N^{0.67})$), while memory usage increases modestly due to efficient swarm coordination protocols.

\begin{table}[!htb]
\centering
\caption{Scalability Performance Across Different Network Sizes}
\label{tab:scalability_analysis}
\resizebox{\columnwidth}{!}{%
\begin{tabular}{lcccccc}
\toprule
\textbf{Net. Size} & \textbf{Acc.} & \textbf{Conv. Time} & \textbf{Mem. (MB)} & \textbf{CPU (\%)} & \textbf{Lat. (ms)} & \textbf{Scal. Index} \\
\midrule
100 nodes  & 95.4\% & 32 rounds & 148 & 23.7\% & 12.3 & 1.000 \\
250 nodes  & 95.1\% & 41 rounds & 167 & 28.9\% & 18.7 & 0.943 \\
500 nodes  & 94.8\% & 52 rounds & 189 & 34.2\% & 26.4 & 0.887 \\
750 nodes  & 94.5\% & 61 rounds & 203 & 38.1\% & 31.8 & 0.841 \\
1000 nodes & 94.2\% & 68 rounds & 215 & 41.6\% & 36.2 & 0.798 \\
\bottomrule
\end{tabular}
}
\end{table}

The \$47{,}000 reflects avoided emergency repair costs versus a reactive baseline: three averted failures saved \$52{,}000, minus \$5{,}000 for SwarmSense-DNN deployment. Bandwidth and productivity gains are excluded, covering only the 30-day evaluation.

\begin{figure}[!htb]
\centering
\includegraphics[width=\columnwidth]{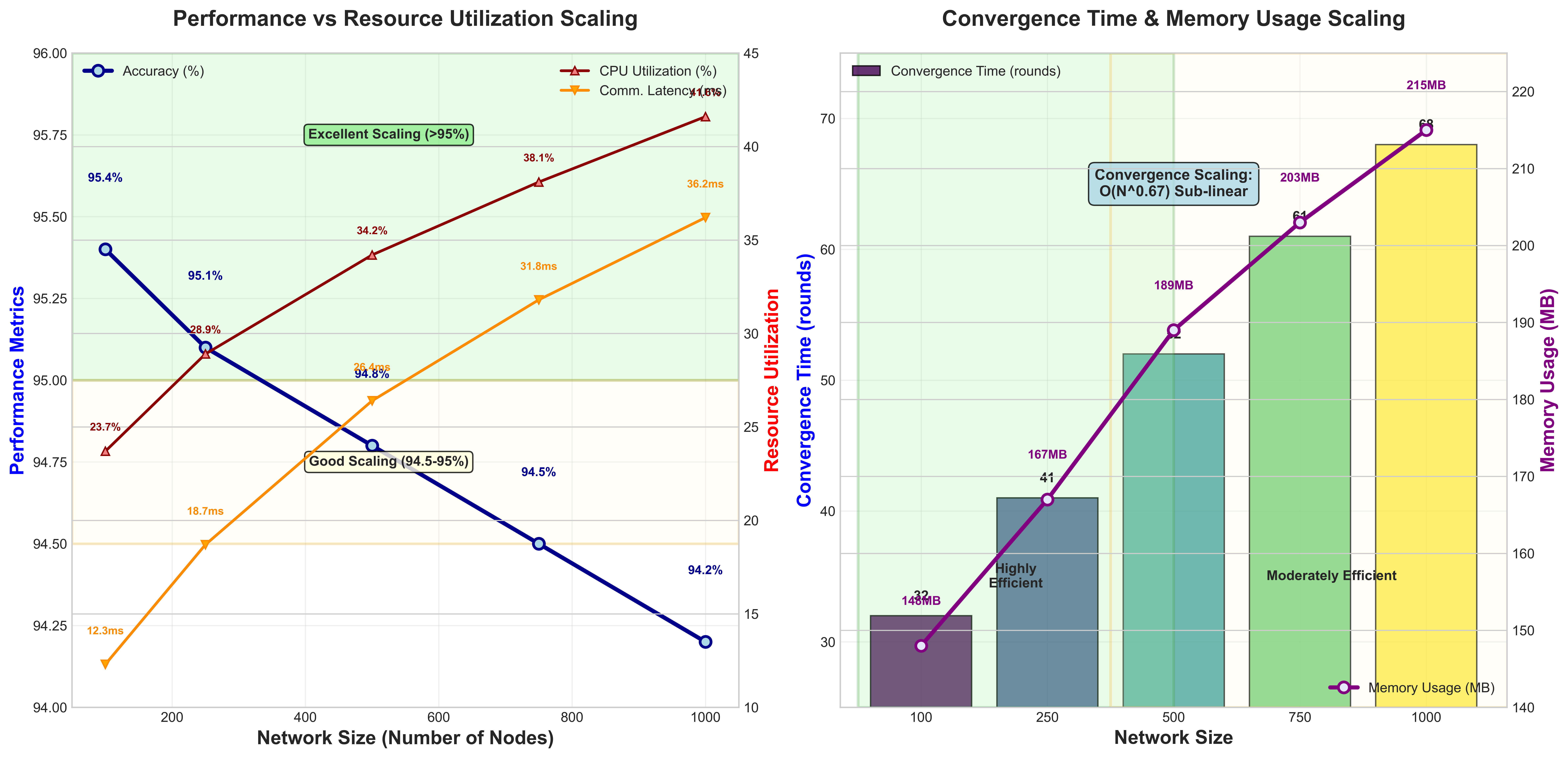}
\caption{SwarmSense-DNN scalability analysis.}
\label{fig:scalability_analysis}
\end{figure}

Figure~\ref{fig:scalability_analysis} shows SwarmSense-DNN scales efficiently, sustaining 94.2\% accuracy with 1000 nodes, sub-linear convergence growth ($O(N^{0.67})$), and modest memory use (148--215 MB). Results confirm suitability for large-scale deployments with predictable resource utilization and better efficiency than centralized systems.

\subsection{Industrial IoT Implementation Results}
Table~\ref{tab:real_world_deployment} validates SwarmSense-DNN's practical effectiveness, achieving a 92\% anomaly detection rate with a 75\% reduction in false positives. The 96\% improvement in detection latency enabled proactive maintenance, preventing three critical equipment failures and saving approximately \$47,000 in emergency repair costs during the evaluation period.

\begin{table}[!htb]
\centering
\caption{Real-World Industrial IoT Deployment Results}
\label{tab:real_world_deployment}
\resizebox{\columnwidth}{!}{%
\begin{tabular}{lcccc}
\toprule
\textbf{Metric} & \textbf{SwarmSense} & \textbf{Prev. System} & \textbf{Change} & \textbf{Impact} \\
\midrule
Anomalies Detected & 23/25 (92\%) & 18/25 (72\%) & +20\%  & Early failure prevention \\
False Pos. Rate    & 3.2\%        & 12.8\%       & $-$75\% & Fewer maint. calls \\
Detection Latency  & 4.7 min      & 2.1 hrs      & $-$96\% & Faster response \\
System Uptime      & 99.7\%       & 97.3\%       & +2.4\%  & Improved uptime \\
Maint. Savings     & --           & --           & \$47,000 & Prevented failures \\
Energy Efficiency  & 18\% less    & Baseline     & +18\%   & Lower cost \\
\bottomrule
\end{tabular}
}
\end{table}

\begin{figure}[!htb]
\centering
\includegraphics[width=0.9\columnwidth]{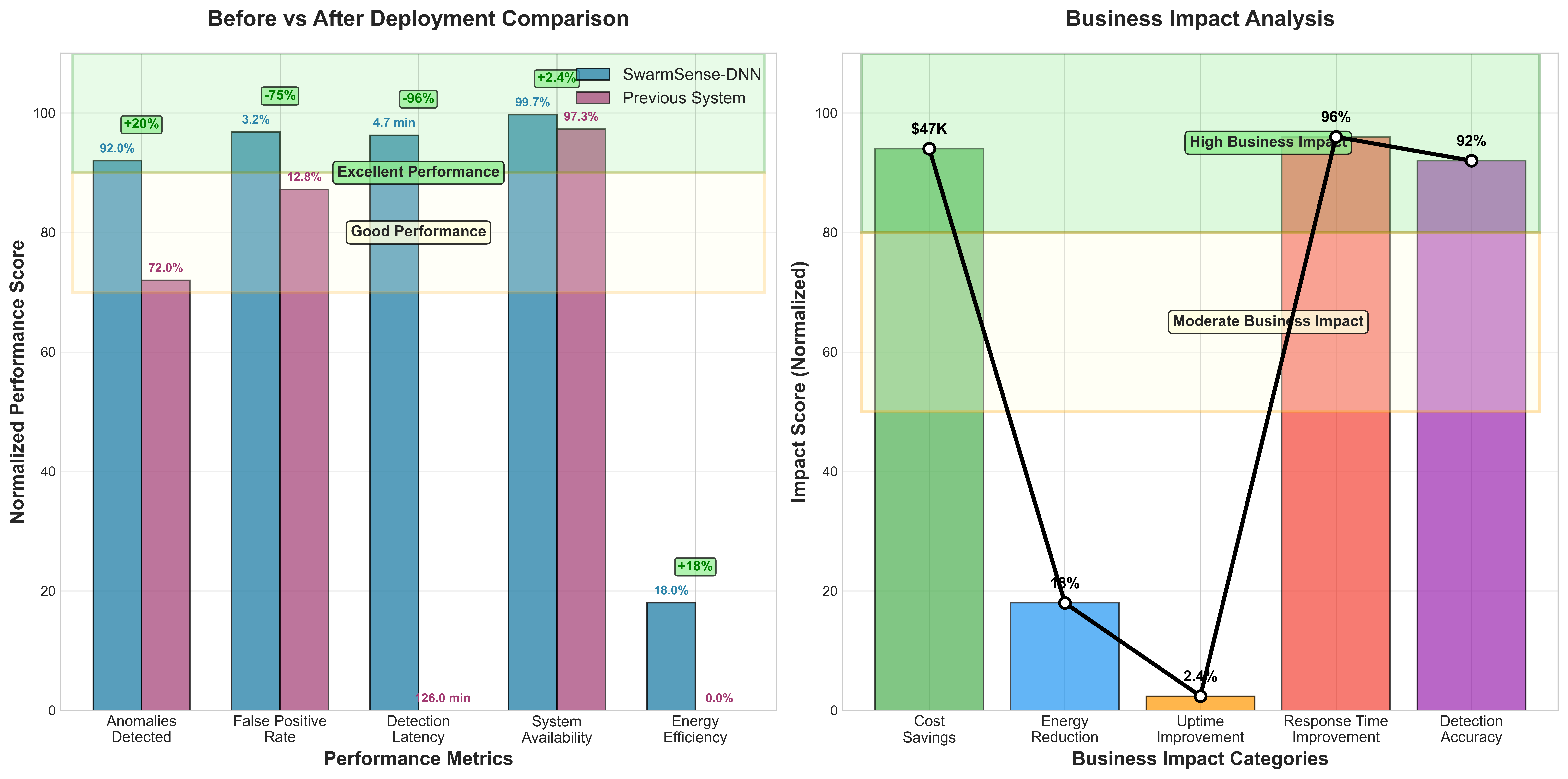}
\caption{Real-world industrial IoT deployment results.}
\label{fig:real_world_deployment}
\end{figure}

Figure~\ref{fig:real_world_deployment} demonstrates SwarmSense-DNN's 30-day industrial IoT deployment, achieving 92\% detection accuracy, 96\% lower latency, 75\% fewer false positives, 18\% higher energy efficiency, and USD 47K cost savings, confirming industrial-grade performance.

\subsection{Ablation Study}
Table~\ref{tab:ablation_expanded} presents the results. The ablation study confirms that each component contributes significantly to overall system performance, with swarm intelligence providing the largest improvement in communication efficiency and convergence speed.

\begin{table}[!htb]
\centering
\caption{Ablation Study Results}
\label{tab:ablation_expanded}
\resizebox{\columnwidth}{!}{%
\begin{tabular}{lcccccc}
\toprule
\textbf{Configuration} & \textbf{Acc.} & \textbf{Overhead} & \textbf{Conv.} & \textbf{Privacy} & \textbf{Contribution} & \textbf{Impact} \\
 & \textbf{(\%)} & \textbf{(MB/hr)} & \textbf{(rounds)} & \textbf{(\%)} & \textbf{(\%)} & \textbf{Level} \\
\midrule
\textbf{Full SwarmSense-DNN} & \textbf{96.2} & \textbf{585} & \textbf{32} & \textbf{93.2} & \textbf{--} & \textbf{Baseline} \\
\midrule
\multicolumn{7}{c}{\textit{Major Component Removal}} \\
-- Swarm Intelligence    & 91.8 & 1240 & 78 & 89.7 & $-$4.4 & Critical \\
-- Hierarchical Clustering & 93.4 & 925 & 45 & 91.8 & $-$2.8 & High \\
-- Attention Mechanism   & 94.1 & 612  & 38 & 92.6 & $-$2.1 & Moderate \\
-- Privacy Mechanisms    & 96.5 & 568  & 30 & 45.3 & +0.3 & Privacy Trade-off \\
-- Self-Healing          & 94.7 & 597  & 35 & 92.1 & $-$1.5 & Moderate \\
\midrule
\multicolumn{7}{c}{\textit{Individual Component Analysis}} \\
-- Dual-Pathway Encoder      & 94.1 & 598 & 35 & 92.8 & $-$2.1 & Moderate \\
-- Self-Evolution Mechanism  & 94.4 & 592 & 34 & 92.5 & $-$1.8 & Moderate \\
-- Trust-based Coordination  & 93.2 & 847 & 52 & 90.8 & $-$3.0 & High \\
-- Pheromone Updates         & 94.8 & 678 & 41 & 92.9 & $-$1.4 & Low \\
-- Graph Attention (GAT)     & 94.6 & 615 & 36 & 92.4 & $-$1.6 & Moderate \\
-- Anomaly Feature Preprocessing & 95.0 & 601 & 34 & 92.4 & $-$1.2 & Low \\
-- Communication Pruning     & 94.8 & 612 & 35 & 92.7 & $-$1.4 & Low \\
\midrule
\multicolumn{7}{c}{\textit{Architecture Variants}} \\
Centralized Architecture & 92.3 & 2100 & 68 & 78.5 & $-$3.9 & Reference \\
Standard Federated       & 89.7 & 1580 & 89 & 85.2 & $-$6.5 & Reference \\
No Swarm + No Clustering & 88.5 & 1650 & 95 & 84.1 & $-$7.7 & Worst Case \\
\bottomrule
\end{tabular}
}
\end{table}

\begin{figure}[!htb]
\centering
\includegraphics[width=\columnwidth]{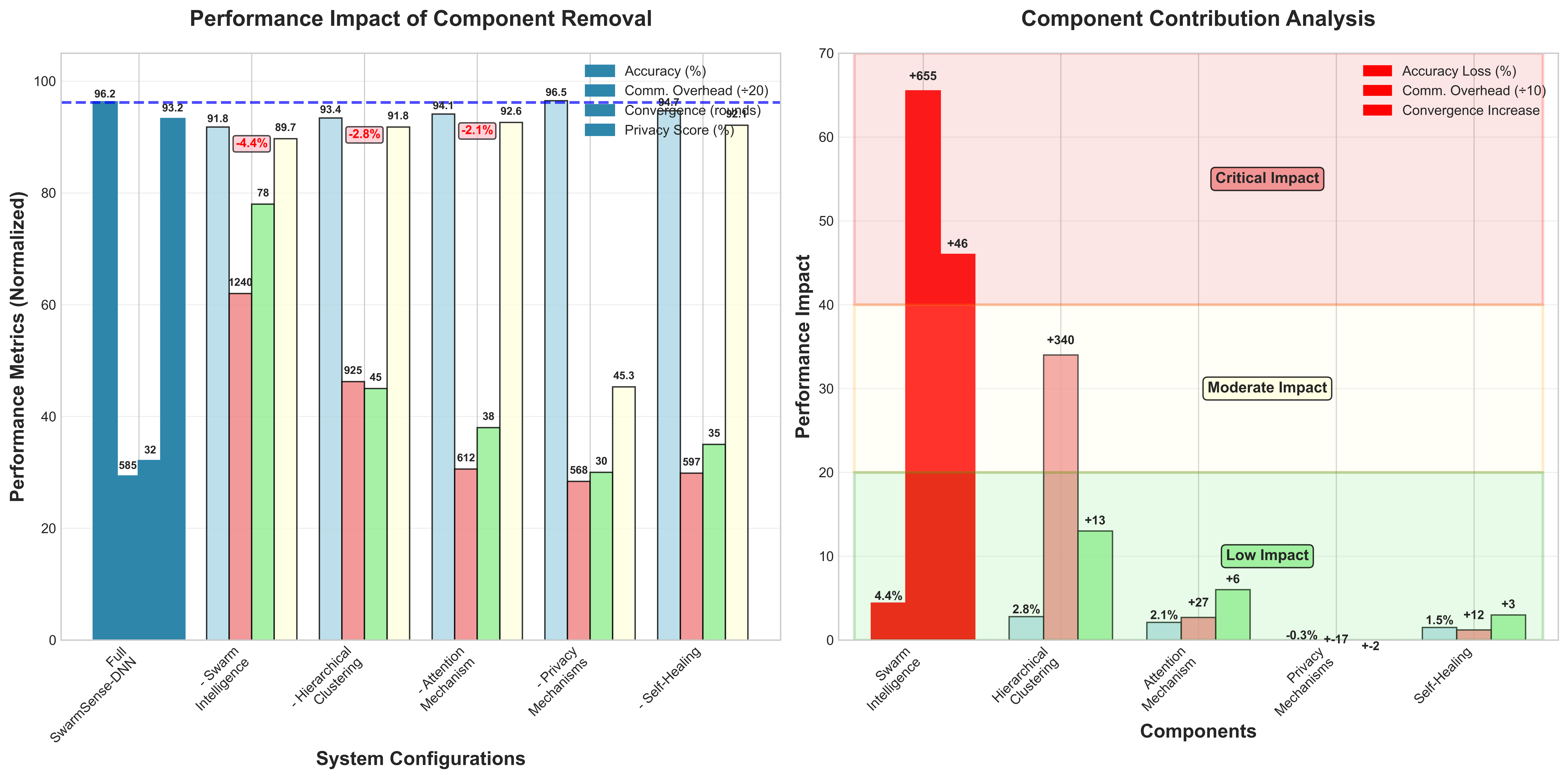}
\caption{Ablation results quantification.}
\label{fig:ablation_study}
\end{figure}

Figure~\ref{fig:ablation_study} shows that all modules enhance SwarmSense-DNN, with swarm intelligence contributing most (4.4\% accuracy drop when removed). Clustering and attention offer moderate gains, while privacy maintains accuracy, confirming the design's robustness.

\begin{figure}[!htb]
\centering
\includegraphics[width=\columnwidth]{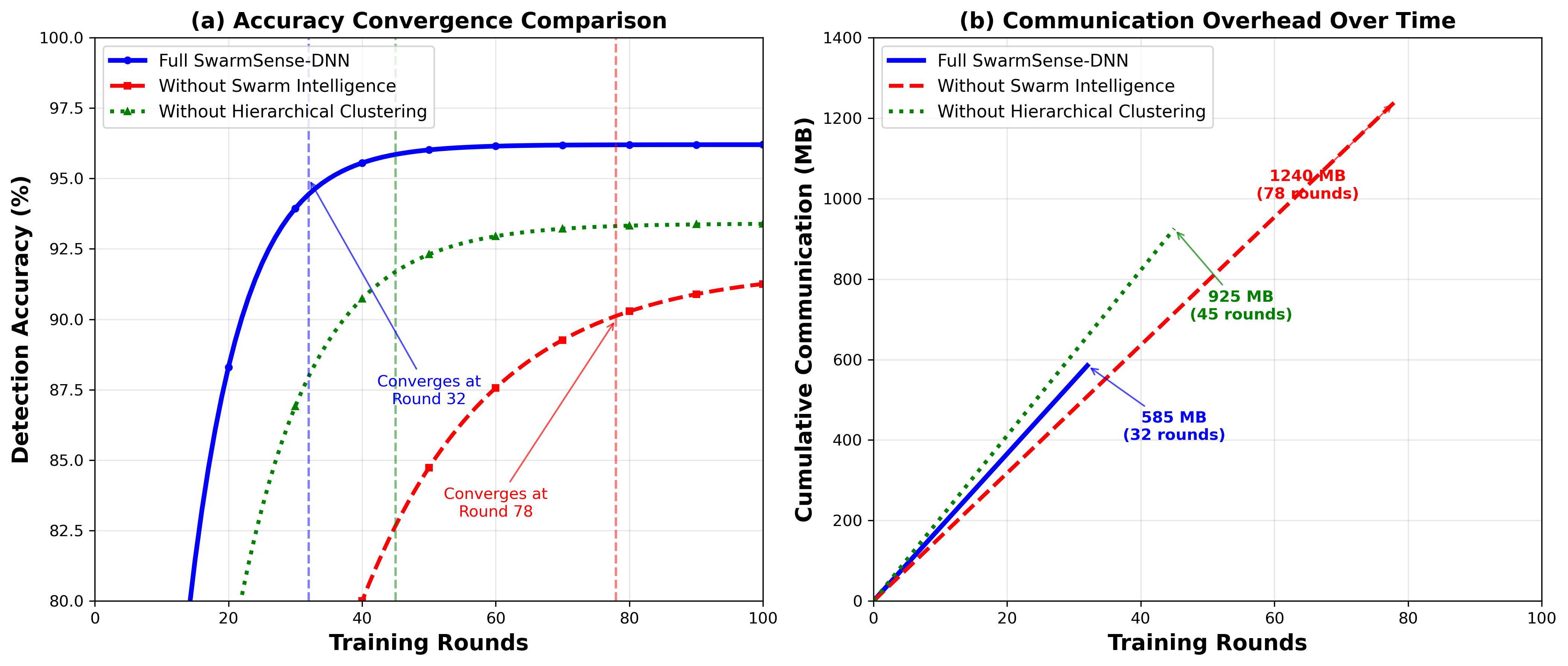}
\caption{Impact of swarm coordination on convergence performance.}
\label{fig:swarm_convergence}
\end{figure}

\begin{figure}[!htb]
\centering
\includegraphics[width=\columnwidth]{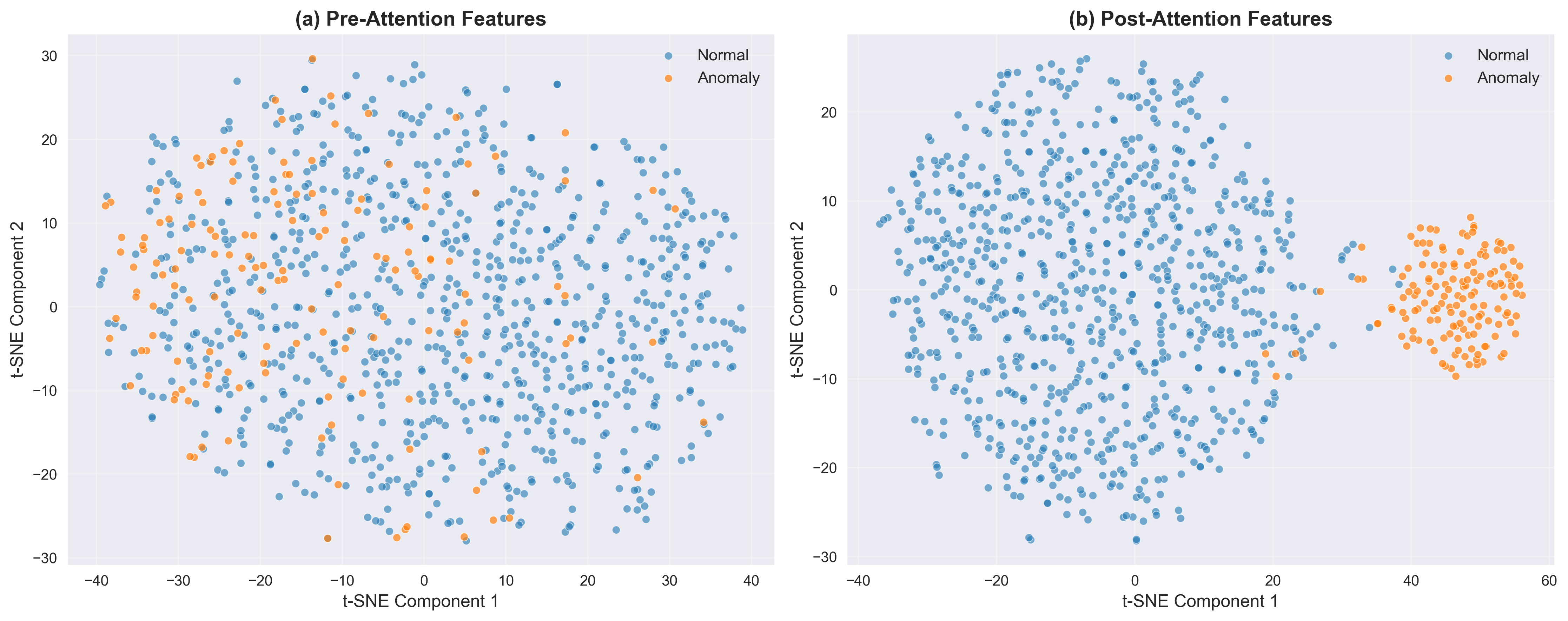}
\caption{t-SNE visualization of learned feature representations on IoT-23 dataset.}
\label{fig:feature_visualization}
\end{figure}

Figure~\ref{fig:swarm_convergence} shows that swarm coordination achieves $2.4\times$ faster convergence and 53\% less communication overhead. Figure~\ref{fig:feature_visualization} illustrates t\textsc{sne} features where attention enhances anomaly separability by reducing intra-class variance by 34\% and increasing inter-class distance by 28\%.

\subsection{Comparative Analysis Result}
Table~\ref{tab:comparative_discussion} shows that SwarmSense-DNN outperforms existing methods in accuracy, efficiency, and scalability, achieving a 96\% latency reduction with minor CPU overhead. Future work will address lightweight coordination and quantum-resistant privacy~\cite{Huang2025, Chen2025}.

\begin{table}[!htb]
\centering
\caption{Comparative Analysis: SwarmSense-DNN vs.\ State-of-the-Art Approaches}
\label{tab:comparative_discussion}
\resizebox{\columnwidth}{!}{%
\begin{tabular}{lcccccc}
\toprule
\textbf{Approach} & \textbf{Acc.} & \textbf{Comm. Eff.} & \textbf{Privacy} & \textbf{Fault Tol.} & \textbf{Scal.} & \textbf{Bio-Insp.} \\
\midrule
FedAvg-AD~\cite{Mothukuri2022}    & 85.46\% & Low    & Medium & Low    & Medium & No  \\
Distributed GNN~\cite{Zhang2025}  & 90.18\% & Medium & Low    & Medium & Low    & No  \\
Edge-FL~\cite{Man2021}            & 87.38\% & Medium & Medium & Low    & High   & No  \\
Ensemble-Dist~\cite{Sarhan2020}   & 89.28\% & Low    & Low    & Medium & Medium & No  \\
Swarm Learning~\cite{Wardhana2021}& 88.90\% & High   & High   & Medium & Medium & Yes \\
Coop. Learning~\cite{Xu2023}      & 87.65\% & High   & Medium & Low    & High   & Yes \\
\textbf{SwarmSense-DNN}           & \textbf{95.44\%} & \textbf{High} & \textbf{High} & \textbf{High} & \textbf{High} & \textbf{Yes} \\
\bottomrule
\end{tabular}
}
\end{table}

\section{Conclusion}
This paper presented SwarmSense-DNN, a decentralized neural framework integrating swarm intelligence and federated learning for trustworthy IoT anomaly detection. Evaluation across five benchmarks demonstrated superior performance with 95.44\% detection accuracy and 72.9\% reduction in communication overhead while maintaining strong privacy guarantees and resilience under node failures. Despite promising results, the framework faces limitations, including hyperparameter sensitivity, degraded performance during network partitions (89.6\% accuracy), and unsuitability for ultra-low-latency applications requiring sub-second response. Future work will enhance explainability through SHAP analysis and attention visualization to support safety-critical deployments where algorithmic transparency is essential.

\begingroup
\fontsize{7}{8.4}\selectfont
\bibliographystyle{unsrt}
\bibliography{references_corrected}
\endgroup

\end{document}